\documentclass[manuscript,screen,acmsmall]{acmart}
\AtBeginDocument{%
  \providecommand\BibTeX{{%
    \normalfont B\kern-0.5em{\scshape i\kern-0.25em b}\kern-0.8em\TeX}}}

\setcopyright{acmcopyright}
\copyrightyear{2025}
\acmYear{2025}
\acmDOI{XXXXXXX.XXXXXXX}

\acmConference[CSCW '25]{The 28th ACM SIGCHI Conference on Computer-Supported Cooperative Work \& Social Computing (CSCW)}{October/November 2025}{unknown}
\acmPrice{15.00}
\acmISBN{}

\usepackage{multirow}
\usepackage{colortbl}
\usepackage{makecell}
\usepackage{csquotes}
\usepackage{array}
\usepackage{CJKutf8}
\usepackage{graphicx}
\usepackage{steinmetz}
\usepackage{booktabs}

\usepackage{amssymb}
\usepackage{xcolor}
\usepackage{tikz}
\usepackage{tabularx}

\newcolumntype{C}[1]{>{\PreserveBackslash\centering}p{#1}}
\newcolumntype{R}[1]{>{\PreserveBackslash\raggedleft}p{#1}}
\newcolumntype{L}[1]{>{\PreserveBackslash\raggedright}p{#1}}

\def\markup{0}

\if\markup1
\usepackage[normalem]{ulem}
  \definecolor{myblue}{rgb}{0,0,0.75}
  \newcommand{\rv}[1]{{\leavevmode\color{myblue}#1}}
  \newcommand{\st}[1]{{\sout{#1}}}
\else
  \newcommand{\rv}[1]{#1}
\newcommand{\st}[1]{}
\newcommand{\sout}[1]{}
\fi

\begin{document}


\title[Using Virtual Reality as a Social Platform]{Systematic Literature Review of Using Virtual Reality as a Social Platform in HCI Community}



\author{Xiaoying Wei}
\orcid{0000-0003-3837-2638}
\affiliation{
  \institution{The Hong Kong University of Science and Technology}
  \city{Hong Kong SAR}
  \country{China}
}
\email{xweias@connect.ust.hk}

\author{Xiaofu Jin}
\orcid{0000-0003-3837-2638}
\affiliation{
  \institution{The Hong Kong University of Science and Technology}
  \city{Hong Kong SAR}
  \country{China}
}
\email{xjinao@connect.ust.hk}

\author{Ge Lin KAN}
\orcid{0000-0002-4422-9531}
\email{gelin@hkust-gz.edu.cn}
\affiliation{
  \institution{The Hong Kong University of Science and Technology (Guangzhou)}
  \city{Guangzhou}
  \country{China}
  }

\author{Yukang Yan}
\orcid{0000-0001-7515-3755}
\affiliation{
  \institution{University of Rochester}
  \city{Rochester}
  \country{USA}
}
\email{}

\author{Mingming Fan}
\authornote{Corresponding Author}
\orcid{0000-0002-0356-4712}
\affiliation{
  \institution{The Hong Kong University of Science and Technology (Guangzhou)}
  \country{China}
}
\email{mingmingfan@ust.hk}

\renewcommand{\shortauthors}{Wei et. al.}

\begin{abstract}

Virtual reality (VR) is increasingly used as a social platform for users to interact and build connections with one another in an immersive virtual environment. Reflecting on the empirical progress in this area of study, a comprehensive review of how VR could be used to support social interaction is required to consolidate existing practices and identify research gaps to inspire future studies.
In this work, we conducted a systematic review of 94 publications in the HCI field to examine how VR is designed and evaluated for social purposes.
We found that VR influences social interaction through self-representation, interpersonal interactions, and interaction environments.
We summarized four positive effects of using VR for socializing, which are relaxation, engagement, intimacy, and accessibility, and showed that it could also negatively affect user social experiences by intensifying harassment experiences and amplifying privacy concerns.
We introduce an evaluation framework that outlines the key aspects of social experience: intrapersonal, interpersonal, and interaction experiences. 
According to the results, we uncover several research gaps and propose future directions for designing and developing VR to enhance social experience.

\end{abstract}




\maketitle

\section{Introduction}

As an emerging communication tool, Virtual Reality (VR) is drawing an increasing number of users into its virtual spaces to interact with one another~\cite{freeman2022working, sancho2009multi, baker2019interrogating, schroeder1997networked}. Unlike traditional visual-audio communication tools, VR delivers a fully immersive experience that simulates face-to-face communication, enabling various interaction cues (e.g., interpersonal distance, body orientation, etc.) to exchange ideas and intentions beyond simply looking at a screen~\cite{Maloney2020, Tanenbaum2020, freeman2022working}. 
This distinct capacity of VR to support social interactions has led to the flourishing of social VR applications in the commercial market \sout{(e.g.,} \rv{such as} Rec Room, BigScreen, AltspaceVR, and VRChat~\cite{freeman2022working, mcveigh2018s}. It has also garnered substantial academic interest in the HCI field, prompting significant research into its application for social purposes. \sout{such as} \rv{This research includes} investigating users' challenges and needs or proposing innovative designs to support social interactions~\cite{sykownik2021most, mcveigh2019shaping, wei2022communication, salminen2018bio, baker2021school, roth2018beyond}.

While existing literature reviews have investigated multi-user interactions within VR environments, their primary focus has been on VR's potential to enhance communication efficiency~\cite{wei2022communication, pan2018and, rubio2017immersive, almoqbel2022metaverse}. 
Although informative, these studies offer limited information on VR's role in fostering emotional connections and nurturing interpersonal relationships, which are crucial for achieving satisfactory social experiences~\cite{devito2007interpersonal, sabri2023challenges, argyle2017social}. 
Recognizing this gap, our systematic literature review aims to understand the use of VR for social purposes from multiple perspectives.

First, it is vital to comprehensively understand the specific VR features that support social interaction and their impact on user perceptions. 
Gaining such insights is crucial for future researchers and designers, as it enables them to customize VR features to meet specific user requirements effectively. 
Although previous research has explored different VR features for social interaction, such as creating expressive avatars to enhance emotional expression~\cite{bernal2017emotional, baloup2021non}, or augmenting user behavior to foster social engagement~\cite{roth2018beyond, dey2017effects},
there is a lack of cohesive analysis regarding the diverse intentions behind these VR feature designs. Therefore, we propose our first research question (\textbf{RQ1}): \textbf{What VR features have been investigated and what are their roles in supporting social interaction?}

Second, gaining a thorough understanding of VR's effects on social interactions and their potential causes is crucial. 
While numerous studies demonstrated VR's benefits including enhancing emotional well-being and offering innovative entertainment experiences~\cite{freeman2021hugging, afifi2022using, zamanifard2023social}, several works also reveal its potential drawbacks, such as leading to lower emotional stability~\cite{lavoie2021virtual, de2019personality} and developmental impacts on young users~\cite{maloney2020complicated, maloney2021stay}. 
Therefore, a comprehensive analysis of VR's effects on social interaction is imperative to provide future researchers with a nuanced understanding of VR's complex influence on social dynamics. 
This knowledge is also crucial in developing VR systems that not only enhance positive experiences but also mitigate negative ones. 
Consequently, we explore \textbf{RQ2: What are the positive and negative effects of using VR for social interaction, and what are their potential causes?}

Last, establishing a structured evaluation methodology is crucial for guiding future researchers in assessing users' social experiences in VR and determining the effectiveness of VR designs. 
However, the evaluation process is complex, requiring the examination of not only communication quality and efficiency but also emotional experiences and interpersonal connections~\cite{devito2007interpersonal, sabri2023challenges, maloney2020anonymity, mcveigh2019shaping}. Furthermore, different studies have investigated various dimensions, employed different settings, and tested with participants of diverse characteristics in their evaluations, contributing to a fragmented understanding of how to evaluate social experiences in VR. For instance, applying inconsistent scales or questionnaires on the same metrics hinders effective comparison and generalization of findings, while variations in experimental settings and participant characteristics affect study outcomes and interpretations.
To summarize current evaluation methods and offer a cohesive framework for future studies, we propose to answer our \textbf{RQ3: How are social experiences in VR evaluated, considering the evaluation dimensions, experimental settings, and participant characteristics?}

We conducted a systematic review following the Preferred Reporting Items for Systematic Reviews and Meta-Analyses (PRISMA) method~\cite{moher2009preferred} to answer these RQs. We identified papers from high-impact venues within the HCI field according to Google Scholar Metrics~\cite{googlemetrics}. 
To include the papers that offer sufficient insights for addressing our RQs, we collected papers based on the following criteria: 1) the paper should explore human-to-human interactions within immersive VR environments, 2) the paper should detail the influence of VR on social experiences, characterized as the range of emotional sensations and responses elicited in users during interpersonal interactions and relationships, including but not limited to aspects like mutual awareness, emotional connections, and affective interdependence~\cite{rime2007interpersonal, salminen2018bio, lee2022understanding, friston2021ubiq, dey2017effects, choudhary2021revisiting}. We finally included a total of 94 articles for further analysis and investigation.

Our study offers multiple contributions to the HCI community. 
First, we find that VR influences social interaction in three key ways, including leveraging self-representation to shape users' self-perceptions and social experiences; offering various interaction strategies to create a natural and engaging experience; and providing interaction environments that set the context and norms to scaffold social behaviors.
Second, we show that while VR offers numerous benefits for social interaction, including promoting relaxation, enhancing engagement, fostering intimacy, and improving accessibility, it also presents challenges like harassment and privacy issues that require future attention and resolution. 
Third, we summarize the evaluation methodology employed in the reviewed studies, considering evaluation dimensions, experimental settings, and participant characteristics, to unify evaluation methods and provide a cohesive framework for future studies.
Fourth, based on our findings, we discuss research directions that can inform future studies in social VR. Our goal is to outline open issues in current studies, to develop a more positive and harmonious social atmosphere in VR, and to reflect on the ecological validity and rigorousness of the existing empirical work.

\section{Related work}

\subsection{Evolution of VR for Social Interaction}

VR is a technology that uses computer-generated simulations to create a three-dimensional environment that users can experience and interact with through specialized devices like headsets. Emerging in the 1960s, the initial VR systems were predominantly tailored for individual usage, serving specific sectors such as medicine, flight simulation, automobile design, and military training~\cite{earlyVR}. 
With technology \sout{advanced} \rv{advancing}, VR systems began to facilitate multi-user experiences, enabling users to enter a communal virtual space where they could interact with one another~\cite{Virtuality}. From the late 1990s to the early 21st century, the focus on multi-user VR was mainly channeled into various professional and specialized domains, such as workplace training~\cite{warburton20103d, berge2008multi}, healthcare~\cite{baram2006virtual}, education~\cite{ibanez2011design, dieterle2009multi}, exhibition~\cite{snibbe2009social}, and engineering~\cite{sancho2009multi}. In these works, researchers investigated VR to support successful collaboration by designing effective collaborative systems and providing efficient communication strategies.

The focus on VR as a medium for social purposes has gained substantial attention post-2010. This can be attributed to key technological advancements in VR display and computational capacities~\cite{introduceVR}. Market introduction of accessible and affordable devices like Oculus Rift and HTC Vive \sout{makes} \rv{made} VR become more palatable for people's daily consumption. As hardware constraints have eased, there has been a proliferation of commercial social VR platforms such as Rec Room, BigScreen, AltspaceVR, and VRChat. These platforms offer a compelling alternative to traditional audio-visual communication tools (e.g., phone calls or video chat) in providing a greater sense of co-presence~\cite{smith2018communication, abdullah2021videoconference}, natural and intuitive communication cues~\cite{Maloney2020, Tanenbaum2020, freeman2022working}, and various social activities~\cite{mcveigh2019shaping, smith2018communication, li2019measuring}. These advantages \sout{attract} \rv{have attracted} more and more people to interact, socialize, and build connections in a shared VR environment~\cite{sykownik2021most, freeman2021hugging}.

The emergence of social VR has captured the attention of researchers in the HCI and CSCW fields to explore how to leverage VR to support social interaction. These studies can be categorized into several key areas:
some focus on improving the efficiency of communication by enabling real-time reconstruction of facial and body movements~\cite{wei2019vr, shen2022mouth}; others aim to amplify emotional expressions by visualizing users' affective states~\cite{salminen2018bio, baloup2021non, roth2018beyond, bernal2017emotional}; a different set of studies explores fostering social connections by investigating the design requirements for creating meaningful shared activities~\cite{baker2020evaluating, baker2021school, xu2022social, rothe2021social}; and finally, some delve into the design challenges and needs specific to diverse demographics~\cite{freeman2022re, morris2023don, baker2021avatar}.
This research demonstrated that social interaction in VR offers users benefits such as enhanced companionship and social support~\cite{zamanifard2019togetherness, freeman2021hugging, afifi2022using, wei2023bridging}, which encourages researchers and practitioners to design and develop VR social platforms in the future.

However, previous research on the use of VR for social purposes has been fragmented, making it challenging to grasp how VR is effectively used for socialization and its potential effects.
A comprehensive understanding of this body of work is required to pinpoint common practices that enable subsequent researchers to quickly understand and draw inspiration from previous studies.

\subsection{Gaps in Existing Literature Review}
There are existing literature reviews investigating social interaction in VR~\cite{wei2022communication, pan2018and, almoqbel2022metaverse, rubio2017immersive}. A focal point of these reviews is assessing VR's contribution to enhancing communication efficacy. For example, Wei et al. examine the use of nonverbal cues for effective communication in VR settings~\cite{wei2022communication}. 
Pan's review focuses on applying VR in multi-person psychological experiments and describes the challenges that psychologists may encounter in VR, such as embodiment, uncanny valley, simulation sickness, ethics, and experimental design~\cite{pan2018and}. 
However, these reviews offer limited insights into VR's use in social contexts, such as understanding VR's impact on emotional connections and interpersonal relationships, which are essential for satisfactory social experiences~\cite{devito2007interpersonal, argyle2017social}. To address this, we undertake a systematic review aiming at bridging this knowledge gap. 

Our initial investigation reveals several areas that necessitate further study, leading us to develop three distinct yet interrelated RQs. 
Firstly, we plan to examine specific VR features and their roles in supporting social interaction (\textbf{RQ1}), to understand the current research landscape. 
Secondly, we seek to provide a holistic understanding of VR's positive and negative effects and their causes on social interactions (\textbf{RQ2}), to help researchers build better social VR that promotes positive experiences and minimizes negative ones.
Lastly, we investigate how social experiences in VR are measured (\textbf{RQ3}) to develop a consistent and comprehensive methodology for assessing the effectiveness of VR in facilitating social interactions. These three RQs guided our subsequent paper collection and data analysis.

\section{Methodology}

\begin{figure}[]
\includegraphics[width=5 in]{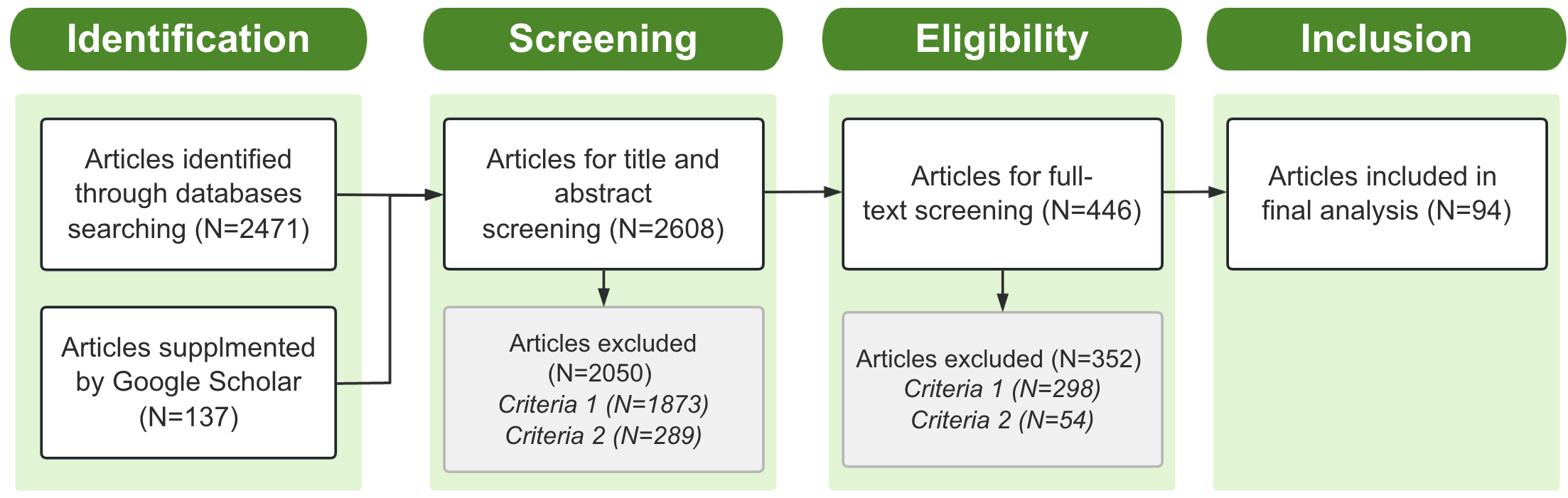}
\vspace{-0.2cm}
\caption{Our literature search and inclusion phases followed the PRISMA procedure. The diagram illustrates the information flow across the four phases: identification, screening, eligibility, and inclusion. It provides the numbers of identified, included, and excluded literature, along with the reasons for exclusions.} 
\Description{This image shows our literature search and inclusion phases following the PRISMA procedure. The diagram illustrates the information flow across the four phases: identification, screening, eligibility, and inclusion. It provides the numbers of identified, included, and excluded literature, along with the reasons for exclusions.} 
\label{fig:process}
\end{figure}

To answer these RQs, we conducted a systematic review of the relevant literature using the PRISMA method~\cite{moher2009preferred}. Following this method, we divided our review process into four phases, which are outlined in Figure~\ref{fig:process}. In the subsequent sections, we provide a detailed explanation of each phase.

\subsection{Identification}

 To identify high-quality research, we used Google Scholar Metrics to select prominent conferences in the fields of HCI and VR~\cite{googlemetrics} to collect papers from, including CHI, CSCW, IMWUT, UIST, IUI, DIS, TOCHI, VRST, TOG, IEEE VR, TVCG, IJHCS, IJHCI, VR, and ISMAR.
 These venues are sourced from five databases: ACM Digital Library, IEEE Xplore, Science Direct, Taylor\&Francis, and Springer.
 Additionally, we conducted a supplementary search using the relevant keywords in Google Scholar to avoid omissions.

 We used the keywords `social*', `interperson*', `multi-user*', and `interact*' to capture articles that discuss social interaction beyond individuals; `virtual reality' and `VR' to ensure that the studies were specifically conducted within the VR medium. We used `social*' to represent variations of the word ``social'' such as `socialize' and `sociality'. The same strategy is applied to `interperson*', `multi-user*', and `interact*'. 
 As each database has its own search logic, we tailored our search queries accordingly. Table~\ref{tab:Boolean} presents the search queries used in five different databases. We allowed these terms to appear in the title or abstract of the articles.

 \begin{table*}[htb!]
    \caption{Boolean instructions for ACM Digital Library, IEEE Xplore, Science Direct, Taylor\&Francis and Springer}
    \Description{The table shows the search queries we used in databases: ACM Digital Library, IEEE Xplore, Science Direct, Taylor\&Francis and Springer}
    \label{tab:Boolean}
    \renewcommand{\arraystretch}{1.3}
    \begin{tabular}{cp{8cm}c}
    \hline 
        \textbf{Database} & \textbf{Boolean Instructions}          \\ \hline
        ACM Digital Library & Title: (social* OR interperson* OR multi-user* OR interact*) AND (``Virtual reality'' OR VR) OR Abstract: (social* OR interpersonal* OR multi-user* OR interact*) AND (``Virtual reality'' OR VR)  \\
        IEEE Xplore & "Abstract": (social* OR interperson* OR multi-user* OR interact*) AND (``Virtual reality'' OR VR)      \\ 
        ScienceDirect  &  Title or Abstract: (social OR socialize OR multi-user OR interpersonal OR interaction OR interact) AND (``Virtual reality'' OR VR) \\
        Taylor\&Francis & Abstract: (social* OR interperson* OR multi-user* OR interact*) AND (``Virtual reality'' OR VR) \\
        Springer  & Title: social* OR interperson* OR multi-user* OR interact* AND ``Virtual reality'' OR VR \\
        Google Scholar &  intitle:((social OR socialize OR multi-user OR interpersonal OR interaction OR interact) AND (``Virtual reality'' OR VR))  \\
        \hline
    \end{tabular}
\end{table*}

We included articles published between January 2013 and June 2023, and written in English. We included full-text papers, works-in-progress, and posters in order to create a comprehensive collection of publications that cover various aspects and applications related to social interaction in VR. Although short papers (e.g., works-in-progress and posters) typically do not have the same expectation of rigor regarding evaluation, we included them because they show the latest explorations of HCI researchers to leverage VR to support social interaction.

The paper identification process resulted in 2608 papers. Specifically, 2471 papers were gathered from five databases. Among the database-sourced papers, 1186 came from the ACM Digital Library, 521 from IEEE Xplore, 638 from Springer, 56 from Taylor\&Francis, and 70 from Science Direct. To ensure comprehensive coverage, we included 137 additional articles from Google Scholar --- we initially selected the top 200 most relevant papers from Google Scholar and, after removing duplicates already presented in the five databases, 137 unique papers remained.

\subsection{Screening and Eligibility}
To select the papers that offer sufficient insights for addressing our RQs, we screened whether the 2608 papers meet the following two criteria:

\begin{enumerate}
\item \textbf{The paper explores multi-user interactions within immersive VR environments.}
Social interactions are the activities or actions that happen between two or more individuals~\cite{shinn1984social}. Therefore, we \textbf{included} articles that investigate interaction strategies or challenges involving at least two users within immersive VR environments. 
In contrast, we \textbf{excluded} articles that only 1) investigate interaction strategies for a single user(e.g., ~\cite{bergstrom2021evaluate, yan2018eyes}); 2) address communication problems between a user and a virtual agent (e.g., ~\cite{pena2020model}); 3) establish connections between immersive VR users and individuals outside the VR environment (e.g., ~\cite{chan2017frontface}); or 4) explore social interaction on 2D virtual world, augmented reality, CAVEs, or other stereoscopic displays (e.g., ~\cite{mack2023towards}).
However, studies investigating from \textit{imagined} partners (e.g., ~\cite{jamil2021emotional, muthukumarana2020touch}) were included as we were interested in the perception of interpersonal interaction in social contexts, even if hypothetical.

\item \textbf{The paper details the influence of specific VR designs on social experiences.}
Our primary objective is to explore how VR is concretely being leveraged for socialization and its effects.
Therefore, we \textbf{included} articles that report and discuss the influence of particular VR features on users' social experiences, in which we characterized ``social experiences'' as the range of emotional sensations and responses elicited in users during interpersonal interactions and relationships, such as mutual awareness, emotional connections, and affective interdependence~\cite{rime2007interpersonal, salminen2018bio, lee2022understanding, dey2017effects, choudhary2021revisiting}.
To be noted, while we acknowledge papers proposing VR designs or systems with potential social interaction applications, we \textbf{excluded} those that did not empirically investigate or detail the impact of specific VR features on social experiences in the paper, such as~\cite{kockro2007collaborative, kan2001internet, qiao2022how, orts2016holoportation}, because these papers may not offer sufficient insight to answer our RQs.





\end{enumerate}

Based on our predefined criteria, all authors independently reviewed the \textit{titles} and \textit{abstracts} of the identified articles during the \textbf{screening} phase. To ensure a consistent understanding of the criteria, the authors conducted five rounds of selection, each with 10 papers. During each round, they discussed their selections and resolved disagreements to reach a consensus. Once researchers established reliability, the remaining articles were evenly divided between the two authors for further screening. This process resulted in the inclusion of 446 articles and the exclusion of 2162 articles.

In the subsequent \textbf{eligibility} phase, we conducted a thorough reading of the full texts of the included articles. The 446 articles were evenly divided between two authors. Each author carefully read and evaluated the full texts against our established criteria. If an article was flagged for removal by one author, it was independently reviewed and verified by another author before being excluded. If a reviewer questioned whether an article met the inclusion criteria, it was marked as ``need to discuss,'' and a final decision was made collectively by all reviewers. Ultimately, 352 articles were excluded for not meeting the criteria, resulting in 94 articles selected for in-depth analysis and investigation.

Among the included articles, 79 were full papers providing detailed information about the experimental methodologies and results, while 15 were extended abstracts (e.g., posters or works-in-progress). These 94 articles are marked with an asterisk (*) in the references.

\subsection{Data Analysis}

We employed \sout{a mixed-methods coding strategy} \rv{an integrated coding strategy} that combined deductive and inductive approaches to analyze the final paper sets. Initially, we employed \textit{deductive method} to categorize findings according to our established RQs, including descriptive statistics of literature (i.e., publication year, venues, research contribution, interaction contexts), VR features influencing social interaction, proven effects of VR usage, and evaluation methodology. These predefined topics provided a structured foundation for our systematic review and analysis. 

Subsequently, we applied \textit{inductive coding} to facilitate the emergence of new themes within the predefined topics, enhancing our flexibility to integrate unforeseen insights and deepen our analysis. In this process, two authors independently coded the same 10 papers from the final paper sets according to the predefined topics and employed Affinity Diagramming to further refine and organize the coding~\cite{beyer1999contextual}. Through collaborative and iterative weekly meetings, they discussed and refined the themes within each topic, incorporating additional insights from advisors to bolster the validity and impartiality of the findings. After reaching a consensus, the two authors independently applied this coding strategy to the remaining literature.
Once all the papers were comprehensively coded, we convened meetings to further refine the final results, ensuring rigor and thoroughness in our analysis.

\section{Results}

\subsection{Overview of the studies and their theoretical foundations}
\label{sec:des}

\subsubsection{Published years and venues}
\begin{figure}[]
\centering
\includegraphics[width=5 in]{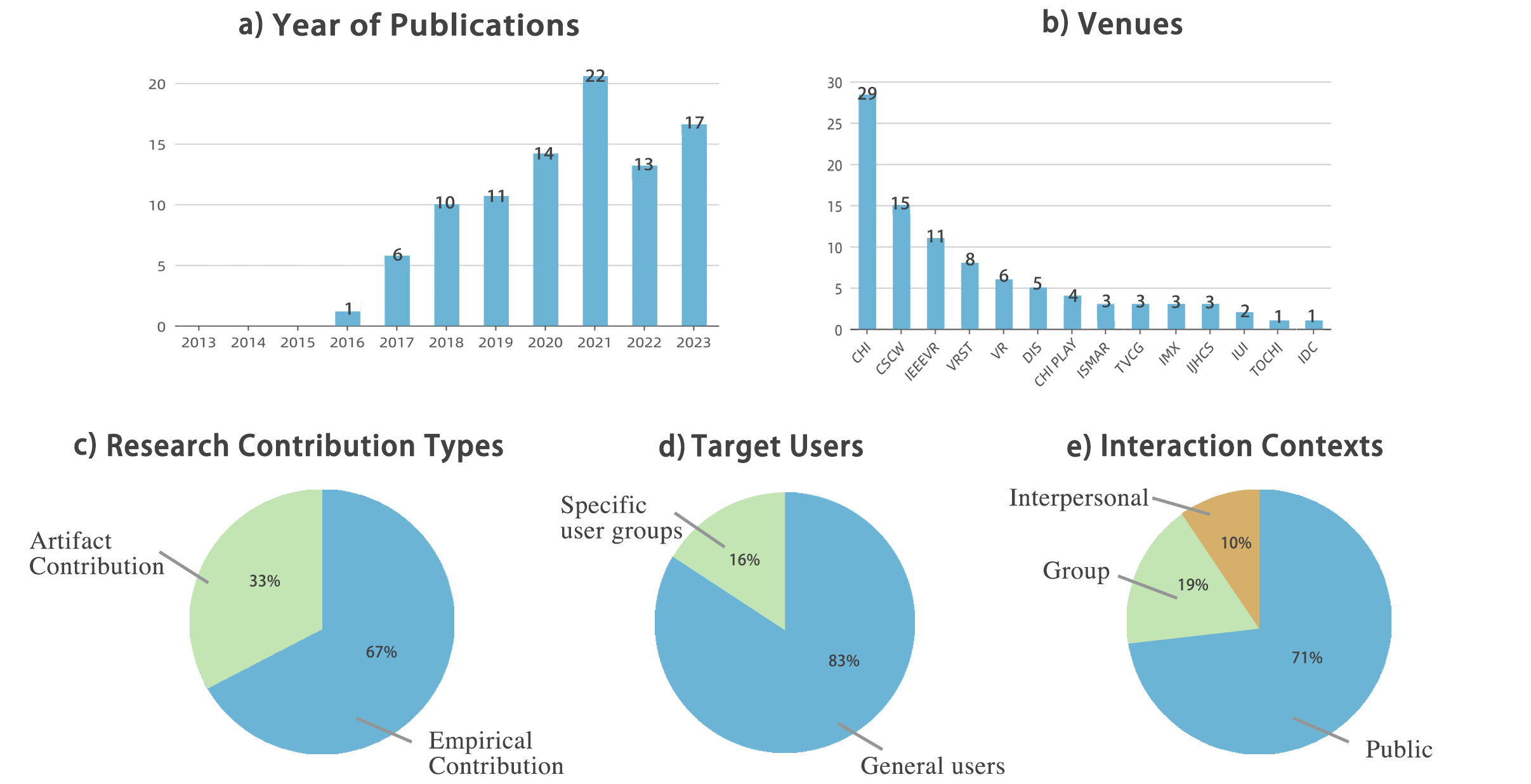}
\vspace{-0.2cm}
\caption{Visualization of descriptive statistics of reviewed paper: (a) shows the years of publication with the number of papers (Please note that the data for 2023 only includes the first six months), (b) shows the venues of publication, (c)(d)(e) show the distribution of research contribution types, target users, and Interaction contexts in the final set of papers} 
\Description{This image shows the visualization of descriptive statistics of the reviewed paper: (a) shows the years of publication with the number of papers (Please note that the data for 2023 only includes the first six months), (b) shows the venues of publication, (c)(d)(e) show the distribution of research contribution types, target users, and interaction contexts in the final set of papers}
\label{fig:general}
\end{figure}

Starting from the year 2016, when VR hardware and software became accessible and affordable for daily use, the number of articles on social experience has gradually increased over the years (as shown in Figure~\ref{fig:general} (a)). Before that, researchers paid more attention to exploring VR for industry proposes (e.g., medical intervention~\cite{parsons2015learning} or construction projects~\cite{porkka2012increased}), than using VR to support daily social interaction. The reviewed articles were published in 14 unique venues as shown in Figure~\ref{fig:general} (b).

\subsubsection{Research contribution types.} Based on the categorization of research contribution types in the HCI field~\cite{wobbrock2016research}, we identified two primary research types among the studies in our paper sets shown in Figure~\ref{fig:general} (c). The majority of the studies (n=63) are empirical contribution research (i.e., "articles that collect, analyze, and interpret observations about known designs, systems, or models, or about abstract theories or subjects''), including interview studies~\cite{freeman2021body, li2023we, maloney2020complicated, baker2019exploring}, quantitative lab experiments~\cite{smith2018communication, abdullah2021videoconference, roth2016avatar, li2021evaluating}, and qualitative field studies~\cite{aburumman2022nonverbal, Tanenbaum2020, Marcel2019towards}, to explore how people interact and communicate in VR to understand their actual needs and challenges. 
The remaining 31 studies contributed valuable artifacts, which focused on designing, building, and evaluating interactive technologies (e.g., system~\cite{salminen2018bio, de2019watching, rothe2021social, montagud2022towards, drey2022towards, xu2022social}, interaction method~\cite{kolesnichenko2019understanding, salminen2018bio, roth2018beyond}, input device/technique~\cite{cha2022performance, wei2019vr}, and hardware toolkit~\cite{muthukumarana2020touch}) that reveal new
possibilities, enable new explorations, facilitate new insights, or compel us to consider new possible futures toward leveraging VR to support users' social interaction. 

\subsubsection{Target users} 
\label{sec:user}
Most studies (n=79) focused on general users (the broader population of users), investigating their general interactions' challenges and proposing interactive technologies to support their VR interaction. Researchers also explored social interaction in more specific groups of users (n=15), including older adults~\cite{xu2023designing, baker2019exploring, baker2019interrogating, baker2021avatar, baker2021school, baker2020evaluating, wei2023bridging, afifi2022using}, children~\cite{maloney2020complicated, fiani2023big}, teenagers~\cite{maloney2021stay}, middle-age women~\cite{morris2023don}, and LGBTQ~\cite{acena2021my, freeman2022re, li2023we, freeman2022queer}, to understand their specific strategies and needs in VR interaction.

\subsubsection{Interaction contexts}
\label{sec:context}
 Social interaction always happens in and is affected by different contexts~\cite{rad2013role, di2019effects}. To understand the research interests in different interaction contexts, we categorize studies into public, group, or interpersonal social contexts~\cite{cathcart1983mediated}. 

 \textit{Public context.} Most reviewed works investigated users' social interactions in public contexts (n=67). These studies aimed to help users better express their ideas and emotions with general users in VR. Most studies leveraged VR features to support social interaction, such as changing avatar appearances~\cite{latoschik2017effect, roth2016avatar, smith2018communication, bernal2017emotional, choudhary2020virtual, fribourg2018studying, lin2023measuring, rzayev2020s, aseeri2021influence, choudhary2020virtual}, reconstructing facial/body movements~\cite{cha2022performance, wei2019vr, vinnikov2017gaze}, or visualizing users' affective states~\cite{lee2022understanding, bernal2017emotional}. Some studies explored the design needs of VR shared activities (e.g., watching movies, dancing, and so on) that encourage users to socially engage with strangers~\cite{sykownik2021most, de2019watching, rothe2021social, li2021evaluating, montagud2022towards, piitulainen2022vibing}. Others investigated ethical risks (e.g., harassment and privacy concerns) that could negatively impact users' social experiences when interacting in public~\cite{sabri2023challenges, wang2021shared, maloney2020anonymity, freeman2022disturbing, shriram2017all, li2023we, blackwell2019harassment}. 

 \textit{Group context.} In group contexts (n=18), researchers mostly focused on improving interaction quality and enhancing group bonding between friends~\cite{sykownik2020like}, alumni~\cite{xu2022social}, team members~\cite{osborne2023being, freeman2022working, abramczuk2023meet, levordashka2023exploration, he2020collabovr}, or a group of people with similar hobbies~\cite{lee2022designing, baker2021school}. Research has explored several activities such as group reminiscence~\cite{baker2019exploring, baker2021school}, social meeting~\cite{osborne2023being, abramczuk2023meet, freeman2022working, wienrich2018social, he2020collabovr}, playing game~\cite{moustafa2018longitudinal, sykownik2020like, huard2022cardsvr, seele2017here}, visiting museum~\cite{roth2018beyond}, attending opera~\cite{lee2022designing}, celebrating graduation~\cite{xu2022social}, group rehearsal~\cite{levordashka2023exploration},   co-design cake~\cite{mei2021cakevr}, and group chatting~\cite{baker2021avatar, baker2019interrogating}. These studies focused on exploring how VR could be designed to support group connections beyond improving users' working efficiency.

 \textit{Interpersonal context.}  Nine studies were explored toward connecting dyads with certain attachments. Such studies mainly investigated how to better cultivate users' emotional bonds and relationships in VR. Two of them explored design considerations of VR systems for dyads with close relationships, such as family relations~\cite{wei2023bridging, afifi2022using}, romantic partners~\cite{zamanifard2019togetherness}, and others~\cite{moustafa2018longitudinal}. The other four systems are tailored to connect pairs using shared activities, such as social meditation~\cite{salminen2018bio}, photo sharing~\cite{li2019measuring},   playing cards~\cite{huard2022cardsvr}, experience sharing~\cite{wang2020again}, and two-player VR games~\cite{dey2017effects}.


\subsubsection{Theoretical foundations of studies}
\label{sec:Theoretical}
 Incorporating established theories into empirical research is crucial, as it offers a solid framework that systematically guides the design and interpretation of research findings. Our review highlights the adoption or development of three main types of theories across the studies we examined. 
 
 Firstly, theories considering social dynamics in 3D real-world settings, like \textit{Personal Space}, have been explored in VR to understand how spatial dimensions and emotional states influence personal boundaries~\cite{williamson2021proxemics, bonsch2018social, choudhary2020virtual}. The researcher used \textit{Similarity Effects}~\cite{montoya2008actual} to investigate the impact of embodied avatar resemblance on persuasion within VR environments~\cite{shih2023do}. Moreover, theories such as \textit{Communication Privacy Management Theory} and \textit{Self-Determination Theory} have been utilized to study information disclosure and enhance user motivation for social interactions in VR, respectively~\cite{wang2021shared, hoeg2021buddy}. Scholars have also applied theories like \textit{Queer Theory} to address social challenges within VR communities, especially regarding marginalized groups~\cite{freeman2022queer}.

 Secondly, researchers have applied and extended the theories originally rooted in screen-based platforms (e.g., online video games) into VR settings. For instance, they adopted theory like the \textit{Proteus Effect}~\cite{yee2007proteus}, initially developed for interactions in online games and web-based chat rooms, to investigate users' perception of themselves when embodied in VR avatars~\cite{shih2023do}. Similarly, concepts from \textit{Nonverbal Communication for Virtual Worlds} have been applied to analyze nonverbal communication in commercial VR applications~\cite{Tanenbaum2020}. 
 
 Thirdly, in addition to the adaptation of theories drawn from non-VR contexts, a few studies have pioneered the development and application of theories specifically tailored for social VR environments. For example, Li and colleagues developed an evaluation framework for measuring photo-sharing experiences in VR~\cite{li2019measuring}, subsequently adopted by multiple studies to assess user experiences in social VR settings~\cite{li2021evaluating, wang2021shared, montagud2022towards, wang2020again}.

 However, despite these advancements, our analysis indicates that a significant portion of studies (82 out of 94) lack substantial theoretical guidance, suggesting a potential oversight in leveraging theory to inform empirical research.

\subsection{VR features and their roles in social interaction (RQ1)}
\label{sec:features}
\begin{figure}[]
\centering
\includegraphics[width=5.6 in]{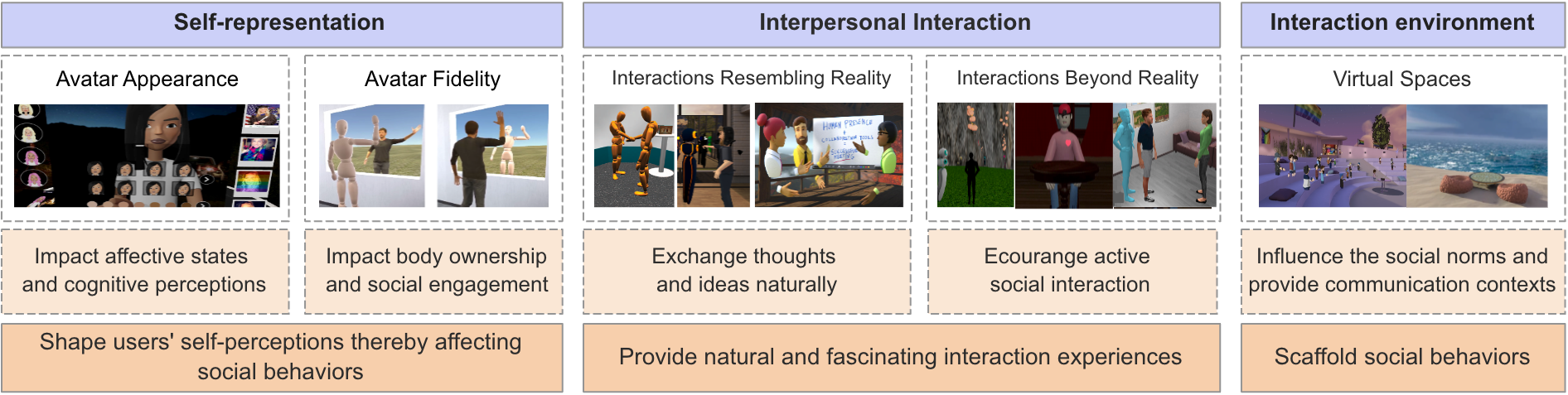}
\caption{This image shows our classifications of VR features, complemented by example figures sourced from reviewed papers: self-representation~\cite{kolesnichenko2019understanding, latoschik2017effect}, interaction strategies~\cite{freeman2022working, Maloney2020, roth2016avatar, lee2022understanding, roth2019technologies}, and interaction environment~\cite{mcveigh2019shaping, li2023we}. We provide the social effect of each VR feature and summarize the social influence of three classifications.} 
\Description{This image shows our classifications of VR features, complemented by example figures sourced from reviewed papers: self-representation, interaction strategies, and interaction environment. This image also shows the social effect of each VR feature and the social influence of three classifications.} 
\label{fig:strategies}
\end{figure}

We categorized our results into \textit{self-representation, interaction strategies}, and \textit{interaction environment}, which refer to how users, the interaction process, and the environment have been investigated to support social interaction, as shown in Fig~\ref{fig:strategies}. 
Within each aspect, we introduce how specific VR features have been investigated in the literature, and their roles in social interaction.

\subsubsection{Self-representation} 
In VR, embodied avatars are digital representations of users~\cite{kolesnichenko2019understanding, mcveigh2019shaping}. This section demonstrates how the visual representation of an avatar shapes users' perceptions and influences their social behaviors. 

First, customizing \textbf{avatars' appearance} enables users to manage their self-image, impacting their self-perceptions and affecting social behaviors. This phenomenon is known as the Proteus Effect~\cite{freeman2021body, yee2007proteus}. In these studies, researchers revealed how different avatars' appearance, such as \textit{aesthetics}~\cite{freeman2021body, wei2023bridging, kolesnichenko2019understanding, bernal2017emotional}, \textit{age}~\cite{morris2023don}, \textit{gender}~\cite{freeman2022re}, \textit{race}~\cite{freeman2022disturbing}, \textit{style}~\cite{lin2023measuring}, \textit{body size}~\cite{choudhary2020virtual, choudhary2021revisiting}, \textit{clothes and accessories}~\cite{piitulainen2022vibing, freeman2022re}, and similarity~\cite{shih2023do}, impact users' affective states and cognitive perceptions during social interaction. For example, embodying a younger avatar can make older people feel more confident, leading to more body movements and interactions with their partners~\cite{wei2023bridging}. 
Additionally, studies also reveal that one's avatar appearance can effectively influence others' willingness to interact~\cite{roth2016avatar, latoschik2017effect, freeman2022re, choudhary2020virtual, wei2022communication, banakou2016virtual, kolesnichenko2019understanding}. For instance, creating a glamorous and well-dressed avatar makes users appear more vibrant and encourages others to engage with them~\cite{freeman2021body}. Therefore, users can customize their avatars' appearance to modify their perceptions and achieve ideal social experiences. 

Studies also investigated the effects of \textbf{avatars' fidelity} (e.g., \textit{visibility}~\cite{latoschik2019not, smith2018communication, aseeri2021influence} and \textit{realism}~\cite{latoschik2017effect, wei2023bridging}) and showed that it could impact users' sense of body ownership (e.g., the feeling that one's body has been substituted by the avatar and that the new body is the source of the sensations~\cite{gonzalez2018avatar}). A higher sense of body ownership users could allow users to immerse themselves in a simulated world and interact with others more authentically and persuasively~\cite{shih2023do}. However, it does not always result in an ideal social experience. A few studies also reveal that the low-fidelity avatars (e.g., cartoonish and unreal) make users feel more relaxed~\cite{wei2023bridging, moustafa2018longitudinal}, offering casual and equal interaction experiences and encouraging active engagement.

\subsubsection{Interactions strategies} 

VR enables various interaction strategies for users to express themselves. In this section, we introduce the dynamic actions and behaviors that users perform within the VR to communicate and interact with others.
We identified two primary interaction strategies discussed in the literature: interactions resembling reality, and interactions beyond reality. 

 \textbf{Interactions resembling reality} are the most common interaction strategies in VR, which provide an intuitive and acceptable way for social interaction since they mimic real-life interactions~\cite{Maloney2020, freeman2022re, moustafa2018longitudinal, wang2020again, smith2018communication}. Among these, \textit{verbal communication} is the most direct way. Features such as pitch, tone, accent, stress, and turn-taking in voice convey information about users' affective states, gender, age, and country of origin, enhancing the richness of communication~\cite{jakobson1972verbal, abdullah2021videoconference, smith2018communication}. To complement verbal communication, users also use a variety of \textit{non-verbal communication cues} to convey their emotions and intentions. For instance, \textit{facial expressions} (e.g., laughter, frowning) and \textit{body movements} (e.g., postures and head orientations) naturally convey feelings such as joy, nervousness, and excitement, adding depth to the social interaction experience~\cite{baloup2021non, Maloney2020, freeman2022working, Tanenbaum2020, yang2022hybridtrak, ahuja2022controllerpose, fang2021social}. 
 \textit{Gaze} indicates attention, regulates intimacy and provides information~\cite{vinnikov2017gaze, seele2017here}, while lip movements improve comprehension of conversations~\cite{yamada2021visual, aseeri2021influence}.
 Additionally, users adjust their \textit{interpersonal distance} in VR based on offline spatial behaviors, nonverbally indicating their level of closeness towards others~\cite{boletsis2017new, Tanenbaum2020, lin2023measuring, choudhary2020virtual, choudhary2021revisiting, williamson2021proxemics}. In addition, users engage in \textit{body contact} through embodied avatars in VR to express the sense of presence or intimacy, such as shaking hands, hugging, and touching~\cite{zamanifard2019togetherness, moustafa2018longitudinal, wei2023bridging, freeman2021hugging, rzayev2020s, sykownik2020like, jamil2021emotional, muthukumarana2020touch, fermoselle2020let}. 

 Additionally, the application of \textit{communication-supportive tools}, such as virtual pens, markers, cards, images, mirrors, and 3D objects, allow users to efficiently express their ideas and thoughts in VR world~\cite{freeman2022working, moustafa2018longitudinal, fu2023mirror}. For instance, users can exchange photos in VR by virtually passing them to one another and highlighting interesting areas to stimulate conversation and discussion about the images~\cite{li2019measuring, schoon2021towards}. Moreover, these communication-supportive tools could also benefit for users to engage in creative and collaborative activities~\cite{freeman2022working, baker2021school, li2019measuring, he2020collabovr}. For example, users could employ virtual markers to sketch their ideas and create 3D illustrations that offer a vivid visualization of concepts, surpassing traditional methods of interaction~\cite{freeman2022working, he2020collabovr}. 


 \textbf{Interactions beyond reality}. As a digital virtual world, VR offers interactions beyond reality for users to express emotional states, enhancing the appeal of social interaction. Reviewed studies showed that the \textit{emotes} (e.g., a smiling face with hearty eyes or star-struck, liking, etc.) serves as an alternative way to convey emotions when the current VR technology may limit the expression of facial expressions and body gestures~\cite{kolesnichenko2019understanding, Tanenbaum2020, baloup2021non}. 
 Reviewed works also explored other virtual effects that can be used in social interaction, such as \textit{particles, creatures, fur, skeuomorphic objects, ambient light, halos}, and \textit{bars}. These effects are employed to visualize users' states~\cite{bernal2017emotional, lee2022understanding, dey2017effects, salminen2018bio} or social behaviors~\cite{roth2018beyond} by changing the size, color, or shape.
 For instance, Salminen's work displayed real-time visualization of brain activation and breathing rate by changing the color of halos around users' avatars to increase affective interdependence and led to better social meditation effects~\cite{salminen2018bio}. 
 Studies found that leveraging virtual effects in VR made social interaction more appealing and attractive, leading to higher valence emotional states and a greater willingness to communicate with others~\cite{moustafa2018longitudinal, wei2022communication, roth2018beyond, salminen2018bio, bernal2017emotional, lee2022understanding, dey2017effects}.

\subsubsection{Interaction environment}
 The review indicates the role of virtual spaces in VR in fostering interactive environments and shaping social experiences. Previous research has demonstrated that the physical settings where communication occurs significantly influence interaction patterns and norms~\cite{devito2007interpersonal}. This phenomenon also applies in the VR world, where users' social norms and interaction behaviors are influenced by the architectural features of virtual spaces~\cite{mcveigh2019shaping}. For instance, the size of the VR environment (e.g., expansive outdoor or confined indoor settings) could impact users' social behavior and interpersonal distances~\cite{sra2018your, williamson2021proxemics}. Consequently, users create various VR spaces and engage in shared activities tailored to their desired social experiences, such as intimate conversations in private rooms~\cite{freeman2022working}, celebratory gatherings in large party rooms~\cite{xu2022social}, or leisurely visits to tourist attractions with family members~\cite{wei2022communication}. Several studies also highlight users' expectations of a customizable and personalized VR environment~\cite{wei2023bridging, mcveigh2019shaping}. 
 \sout{Moreover, the 3D virtual spaces facilitate shared activities in VR, enabling users to engage in ``physical'' interactions despite geographic distances, fostering a sense of togetherness and deep connection~\cite{wei2023bridging, baker2021school}. }
\subsection{Demonstrated effects of VR on social interaction (RQ2)}
\label{sec:effects}

Our literature review and thematic synthesis identified four positive and three negative effects of VR on social interaction. For each identified effect, we present the potential underlying causes.

\subsubsection{Promoting relaxation} 
VR could create relaxing social experiences for users for the following reasons.  
First, VR offers customizable avatars for users to present themselves. This avatar is regarded as a protective shield as it provides a certain level of anonymity for users~\cite{roth2018beyond, wei2023bridging, maloney2020falling, acena2021my, sykownik2022something, sykownik2021most, zamanifard2023social}. Users crafted their preferred self-appearance to interact in the VR world without the fear of judgment or ridicule, which encourages them to initiate communication~\cite{salminen2018bio, freeman2021body, baker2019interrogating, freeman2021hugging}. 
Second, VR provides shared activities that entertain users, allowing them to escape the stresses of daily life and enjoy relaxing experiences together~\cite{maloney2020falling, freeman2021hugging, baker2021avatar, acena2021my, maloney2020complicated, baker2021school, maloney2021stay, sykownik2021most, xu2022social, piitulainen2022vibing}. Additionally, they could alleviate users' concerns about encountering awkward silent moments by offering conducive shared contexts that stimulate conversation topics and maintain a natural conversational flow~\cite{mcveigh2019shaping, wei2023bridging, freeman2021hugging}. 
Last, VR offers an easy-to-escape social environment to provide a safe experience and encourage socialization. For example, users can interact with others in virtual clubs while physically staying in their bedrooms. They could easily escape from the virtual environment when they feel uncomfortable by taking off the headset~\cite{blackwell2019harassment, freeman2022disturbing, li2023we}. This reduces anxiety during VR use~\cite{bosch2013professional, morie2014avatar, waltemate2018impact}. Overall, findings from most reviewed studies confirmed that social interaction in VR could help them relax and alleviate the stresses of everyday life~\cite{maloney2020falling, freeman2021hugging, baker2021avatar, acena2021my, maloney2020complicated, baker2021school, maloney2021stay, sykownik2021most, xu2022social, piitulainen2022vibing}. 

\subsubsection{Enhancing engagement}

VR fosters users' engagement in social interactions for three reasons.
First, immersive VR environments provide a higher sense of co-presence, making users feel like they are actually in the same physical space with others, even if they are geographically separated~\cite{smith2018communication, roth2016avatar, latoschik2017effect, aseeri2021influence, li2021evaluating, li2019measuring}. This feeling of being ``present'' together lets them be deeply engaged, absorbed, and mentally connected with the social interaction~\cite{lee2023so, moustafa2018longitudinal, baker2021avatar, wei2023bridging, freeman2021hugging, rzayev2020s, sykownik2020like}.
Furthermore, the full-body tracking avatar in VR (e.g., one's avatar mirrors their physical body movements in real-time) allows users to intuitively express their thoughts through interactions that replicate the natural flow of conversation (e.g., body language, interpersonal distance) to reduce the cognition burden that may distract engagement~\cite{smith2018communication, abdullah2021videoconference, he2020collabovr}. 
Finally, the VR headset could contribute to users' engagement by blocking out environmental distractions such as noise and visual stimuli from reality (e.g., notifications in smartphones, surrounding people, and so on)~\cite{smith2018communication, li2019measuring, roth2018beyond, salminen2018bio, dey2017effects}. This allows users to prioritize the interaction in the VR world and devote their attention to their partner. However, two studies also highlighted safety concerns brought by VR headsets' limitation of environmental awareness~\cite{abramczuk2023meet, wei2023bridging}. Participants expressed worries about falls and collisions as VR headsets obstruct their view of the physical world, emphasizing the need for safety while enjoying VR benefits~\cite{wei2023bridging}. Future VR designers need to adopt methods to balance users' focus on the VR environment with an awareness of external circumstances for safer VR social interactions, such as integrating distracting stimuli from the physical surroundings into the VR experience~\cite{tao2022integrating}.

\subsubsection{Fostering intimacy}

 VR demonstrates the potential to cultivate intimacy for two reasons.
 On the one hand, VR offers multimodal interactive feedback, encompassing visual and spatial information, tactile feedback, etc.~\cite{sykownik2021most, Tanenbaum2020, Maloney2020}. Coupled with the sense of embodiment provided by the embodied avatar, users can engage in more intimate contact to convey closeness compared to other visual-audio social media~\cite{inkpen2013kids, brubaker2012focusing, abdullah2021videoconference}. For instance, users could approach or even touch close friends in VR~\cite{ahmed2016reach, pamungkas2016electro, hamdan2019springlets}.
 On the other hand, VR overcomes geographical limitations and enables users to engage in VR activities physically. This helps to foster conversation topics and strengthen mutual understanding, which could help to get and build stronger bonds~\cite{wei2023bridging}. This ability is particularly important for remote people who are in close relationships~\cite{zamanifard2019togetherness, wei2023bridging, moustafa2018longitudinal}. For example, geographically dispersed grandparents and grandchildren indicated that entering the virtual world together and interacting with each other help them alleviate their nostalgia~\cite{wei2023bridging}.

\subsubsection{Improving accessibility}

 VR enables social interaction resembling reality while keeping the accessibility and convenience of online social media. 
 Users from different locations could use VR to transcend geographical boundaries and engage within different virtual environments~\cite{freeman2021hugging, mcveigh2019shaping, wei2023bridging}. 
 Studies revealed that the accessibility and convenience of VR interaction are particularly important for people with mobility constraints~\cite{baker2019interrogating, wei2023bridging, baker2021school}. For example, Baker et al.~\cite{baker2021school} developed a bespoke VR application that allows geographically dispersed older adults to meet in a virtual environment and reminisce about their school experiences. This inspires future researchers to explore the application of VR social interactions among people with mobility constraints, such as the elderly in nursing homes and individuals with disabilities facing travel difficulties. This not only ensures their convenience and safety, but also helps them establish closer social connections and gain a degree of social support~\cite{baker2021school, acena2021my, van2023feelings}.

\subsubsection{Intensifying harassment experiences.}

 Harassment was described as any interaction or experience that intentionally upsets them and causes harm, aggravation, anxiety, and instability~\cite{freeman2022disturbing}. 
 Reviewed articles indicated that harassment in VR can be more severe compared to other forms of online social media due to the strong sensation of presence and embodiment created by embodied avatars~\cite{sykownik2020experience, soni2018see}. This heightened immersion exposes users to malicious behaviors, such as unwanted touching, obstructing movement, and throwing objects, in a more realistic manner, increasing their vulnerability compared to less immersive audio-visual social media platforms~\cite{blackwell2019harassment}. Furthermore, synchronous voice chat in VR facilitates the easy spread of offensive content, such as personal insults, hate speech, and sexualized language. Additionally, the anonymity afforded by VR enhances the likelihood of spreading harmful content, as users feel less accountable for their actions~\cite{blackwell2019harassment, fermoselle2020let}. 
 Despite the urgent need to prevent and mitigate harassment in VR, it poses significant challenges for future researchers, including the lack of consensus amongst VR users of ``harassment'', the lack of documentation of harassment, and the lack of unbiased moderators~\cite{schulenberg2023towards, blackwell2019harassment}. Future research should explore more nuanced methods to address harassment in social VR environments.
 
\subsubsection{Amplifying privacy concerns.} 
VR provides a limited but impactful set of identity cues that can inadvertently expose users' private information, such as voice, avatar appearances, and reconstruction of real environments. While appropriate self-disclosure is necessary to establish social connections and foster closeness with others~\cite{sykownik2022something}, unintentional exposure of sensitive personal information can pose privacy risks~\cite{maloney2020anonymity, maloney2021social}. In VR, the user's voice can unintentionally reveal personal information such as dialect, age, or gender~\cite{blackwell2019harassment}. Similarly, avatar appearances based on users' actual looks can also inadvertently expose information. For example, avatars' height can indicate whether a user is a child or an adult~\cite{blackwell2019harassment}. The exposure of such information can lead to potential harassment, such as users being asked or ridiculed if their voice does not match their avatar's gender identity~\cite{acena2021my} or teenagers being mocked for their pre-puberty voice~\cite{maloney2020complicated}. While current studies are striving to achieve more efficient interaction by reconstructing users' personal information (e.g., appearance, body and facial movements, environment, and biosignals) to achieve higher communication quality in the VR environment~\cite{wei2019vr}, these techniques also raise more privacy concerns~\cite{moge2022shared, sykownik2022something}. How to balance the reconstruction of realistic users' identities and the concerns about privacy exposure in VR remains a challenge.

\subsubsection{Feeling inconvenient and uncomfortable.} While studies show that participants were generally positive about VR's possibility to facilitate interaction, they also perceive traditional visual-audio media, such as phone calls or video chats, to be more convenient and practical for their daily communication~\cite{mei2021cakevr, baker2021avatar}. They suggested the \textit{overhead of initiating communication} \sout{in VR} reduces users' willingness to use VR as a communication medium in their daily life~\cite{levordashka2023exploration, baker2021avatar, wei2023bridging, osborne2023being}. Moreover, users encounter challenges interacting with others in VR due to the \textit{non-intuitive nature of hand controllers}, which increases cognitive load and distracts from effective communication~\cite{galais2019natural, wei2023bridging}. In addition to operational difficulties, discomfort issues with VR devices are prevalent. Studies have shown that \textit{the weight of VR devices} can induce user fatigue and serve as constant reminders of the artificial environment~\cite{moustafa2018longitudinal, abramczuk2023meet, li2019measuring}. Furthermore, some users experience \textit{motion sickness} during VR use, impeding their ability to engage actively in social interaction~\cite{abramczuk2023meet}.

\subsection{Measurement of social experience in VR (RQ3)}
\label{sec:measure}

We examine the measurement employed in reviewed studies to measure users' social experiences in VR, focusing on three key aspects: Evaluation Dimensions, Experimental Settings, and Participant Characteristics. By examining these dimensions, our goal is to offer a comprehensive overview of current research practices. Additionally, it could uncover important insights into identifying unresolved issues in evaluation methods, prompting a critical assessment of current study methodologies, and encouraging a more thorough consideration of the ecological validity and methodological rigor in future empirical investigations.

\subsubsection{Evaluation Dimensions}
Researchers evaluated participants' social experiences in VR from different dimensions. Inspired by the taxonomy of social-emotional competencies~\cite{schoon2021towards}, we categorize them into three main aspects: intrapersonal experiences, interpersonal experiences, and interaction experiences, and provide commonly used metrics of reviewed papers for their assessment, as shown in Table~\ref{evaluation}. 
We also provided the comprehensive list in Table~\ref{allmetric} in the appendix to show the measurement tools used in the reviewed paper for reference.

\textbf{Intrapersonal experience.}
Participants' intrapersonal experiences reflect their internal perceptions and ultimately affect their social behaviors~\cite{schoon2021towards}. Researchers have measured participants' intrapersonal experiences by following dimensions.

 \textit{Embodiment} refers to the effect of users partly or fully perceiving a virtual body as their own. A higher sense of embodiment is always accompanied by a high fidelity of avatar and sufficient haptic feedback~\cite{latoschik2017effect, jamil2021emotional, fermoselle2020let}. The sense of embodiment enhances users' feelings of immersion, and behavioral conforms towards avatar appearance, thereby impacting their social behaviors in VR~\cite{latoschik2017effect, wei2023bridging, moustafa2018longitudinal, freeman2021body}. Seven studies measured users' embodiment through users' self-reported data on subjective scales, such as The Illusion of Virtual Body Ownership\cite{roth2017alpha, latoschik2017effect} and Avatar Embodiment Questionnaire~\cite{gonzalez2018avatar}. The example question is ``I felt as if the body I saw in the mirror might be my body''~\cite{latoschik2017effect}.

 \textit{Presence} (also known as telepresence and spatial presence) is users' illusion of being in the virtual environment~\cite{schroeder2001social, nowak2003effect}, which is measured by 23 studies. This perception affects users' engagement with the virtual world and fosters genuine interactions~\cite{witmer1998measuring}. For example, users experiencing lower levels of presence may perceive themselves as mere observers rather than active participants. The measurements primarily relied on users' self-reported data through subjective scales, such as the Slater-Usoh-Steed Questionnaire~\cite{usoh2000using} and the Witmer and Singer Presence Questionnaire~\cite{witmer1998measuring}. The example question is ``To what extent did you feel like you were inside the environment you saw?''~\cite{roth2016avatar}.

 \textit{Affective states} indicate users' emotional and mood-related conditions during social interaction, impacting users' willingness to interact with others either positively or negatively. 14 studies measured affective states by users' self-reported data through subjective questionnaires, such as Pictorial Mood Reporting Instrument~\cite{vastenburg2011pmri} and Positive and Negative Affect Schedule~\cite{watson1988development, mackinnon1999short}. The example question is ``I felt happy/excited/relaxed/tense/irritated...''~\cite{li2019measuring}. Two studies measured this experience by analyzing participants' verbal behaviors such as valence and arousal~\cite{li2019measuring} or the duration of laughter~\cite{sykownik2020like}.

\begin{table}[]
\renewcommand{\arraystretch}{1.5}
\caption{Summary of aspects of user experience affecting social interaction in VR. The table outlines the evaluation dimensions, describes how each experience influences social interaction, identifies contributing VR features, and lists commonly used metrics (with color \textcolor{cyan}{blue}) and tracking data (with color \textcolor{magenta}{red}) for assessment.} 
\Description{Summary of Aspects of User Experience Impacting Social Interaction in VR. The table outlines the evaluation dimensions for intrapersonal, interpersonal, and interaction experiences, citing the number of papers that have explored each aspect. Additionally, it describes how each type of experience influences social interaction in VR, identifies contributing VR features, and lists commonly used metrics and tracking data for assessment. Example papers for each category are also provided for reference.} 
\label{evaluation}
\resizebox{\columnwidth}{!}{%
\begin{tabular}{l|l|l|l|l|l|l}
\toprule[2pt]
 & \multicolumn{1}{c|}{\begin{tabular}[c]{p{1.8cm}}\textbf{Evaluation Dimensions}\end{tabular}} &  \begin{tabular}[c]{p{0.7cm}}\textbf{Paper Count}\end{tabular} & \multicolumn{1}{c|}{\begin{tabular}[c]{p{3.5cm}}\textbf{Influences on}\end{tabular}} & \begin{tabular}[c]{p{3.5cm}}\textbf{Influenced by}\end{tabular} & \multicolumn{1}{c|}{\begin{tabular}[c]{p{4cm}}\textbf{Commonly used scale and tracking data}\end{tabular}} & \begin{tabular}[c]{p{1cm}}\textbf{Example studies} \end{tabular} \\ \toprule[1pt]
\multirow{4}{*}{\textbf{\begin{tabular}[c]{@{}l@{}}Intrapersonal\\ Experiences\end{tabular}}} & \begin{tabular}[c]{p{1.8cm}}Embodiment\end{tabular} & 7 & \begin{tabular}[c]{p{3.5cm}} Heightens immersion, Behavioral conforms to avatar appearance \end{tabular} & \begin{tabular}[c]{p{3.5cm}} Avatar fidelity, Haptic feedback, etc. \end{tabular}  &  \begin{tabular}[c]{p{4cm}} \tikz\draw[cyan,fill=cyan] (0,0) circle (.7ex); The Illusion of Virtual Body Ownership\cite{roth2017alpha}\end{tabular} & ~\cite{latoschik2017effect, fribourg2018studying} \\ \cline{2-7} 
 & \begin{tabular}[c]{p{1.8cm}}Presence\end{tabular} & 23 & \begin{tabular}[c]{p{3.5cm}} Increases commitment and Fosters genuine interactions \end{tabular}  & \begin{tabular}[c]{p{3.5cm}} Realistic spaces, Spatial audio, etc. \end{tabular} & \begin{tabular}[c]{p{4cm}}\tikz\draw[cyan,fill=cyan] (0,0) circle (.7ex); Slater-Usoh-Steed Questionnaire~\cite{usoh2000using}\end{tabular} & ~\cite{ latoschik2019not, drey2022towards} \\ \cline{2-7} 
 & \multirow{2}{*}{\begin{tabular}[c]{p{1.8cm}}Affective states\end{tabular}}& \multirow{2}{*}{14} & \multirow{2}{*}{\begin{tabular}[c]{p{3.5cm}} Impacts users' willingness of social interaction either positively or negatively \end{tabular}}  & \multirow{2}{*}{\begin{tabular}[c]{p{3.5cm}} Perceptions of overall VR experiences \end{tabular}} &  \begin{tabular}[c]{p{4cm}}\tikz\draw[cyan,fill=cyan] (0,0) circle (.7ex); Positive and Negative Affect Schedule~\cite{watson1988development, mackinnon1999short}\end{tabular} & ~\cite{dey2017effects, wienrich2018social} \\
 &  &  &  &  & \begin{tabular}[c]{p{4cm}}\tikz\draw[magenta,fill=magenta] (0,0) circle (.7ex); Valence and Arousal of speech; Duration of laughter\end{tabular} & ~\cite{li2019measuring, sykownik2020like} \\ \midrule[1pt]
\multirow{4}{*}{\textbf{\begin{tabular}[c]{@{}l@{}}Interpersonal\\ Experiences\end{tabular}}} & \begin{tabular}[c]{p{1.8cm}}Mutual awareness\end{tabular} & 17 & \begin{tabular}[c]{p{3.5cm}} Facilitates mutual attention and responsiveness \end{tabular} & \begin{tabular}[c]{p{3.5cm}} Avatar fidelity, Gesture recognition (e.g., eye contacts, Expressions), etc. \end{tabular} &  \begin{tabular}[c]{p{4cm}}\tikz\draw[cyan,fill=cyan] (0,0) circle (.7ex); Nowak and Biocca questionnaire~\cite{biocca2003toward, nowak2003effect}\end{tabular} & ~\cite{smith2018communication, vinnikov2017gaze} \\ \cline{2-7} 
 & \multirow{2}{*}{\begin{tabular}[c]{p{1.8cm}}Psychological involvement\end{tabular}}& \multirow{2}{*}{21} & \multirow{2}{*}{\begin{tabular}[c]{p{3.5cm}} Deepens connection, Enable more meaningful interaction \end{tabular}}  & \multirow{2}{*}{\begin{tabular}[c]{p{3.5cm}} Emotion sharing features,  Activities encouraging experiences sharing, etc. \end{tabular}} &  \begin{tabular}[c]{p{4cm}}\tikz\draw[cyan,fill=cyan] (0,0) circle (.7ex); Social Connectedness Questionnaire~\cite{van2009social}\end{tabular} & ~\cite{wang2020again, montagud2022towards} \\
 &  &  &  & & \begin{tabular}[c]{p{4cm}}\tikz\draw[magenta,fill=magenta] (0,0) circle (.7ex); Interpersonal distances\end{tabular} & ~\cite{buck2019interpersonal, williamson2021proxemics} \\ \cline{2-7} 
 & \begin{tabular}[c]{p{1.8cm}}Affective Interdependence\end{tabular}& 9  & \begin{tabular}[c]{p{3.5cm}} Enable emotional intertwine, Leads empathic behaviors \end{tabular} & \begin{tabular}[c]{p{3.5cm}} Emotion sharing features, Activities requiring emotional goals, etc. \end{tabular} &  \begin{tabular}[c]{p{4cm}}\tikz\draw[cyan,fill=cyan] (0,0) circle (.7ex); Networked Minds Subscales~\cite{biocca2001networked}\end{tabular} & ~\cite{roth2016avatar, montagud2022towards} \\ \midrule[1pt]
\multirow{5}{*}{\textbf{\begin{tabular}[c]{@{}l@{}}Interaction\\ Experiences\end{tabular}}} & \multirow{2}{*}{\begin{tabular}[c]{p{1.8cm}}Quality of interaction\end{tabular}}& \multirow{2}{*}{11} & \multirow{2}{*}{\begin{tabular}[c]{p{3.5cm}} Determines satisfaction, Shapes the long-term viability of VR platforms \end{tabular}}  &  \multirow{2}{*}{\begin{tabular}[c]{p{3.5cm}} Quality and efficiency of information exchange \end{tabular}}  & \begin{tabular}[c]{p{4cm}}\tikz\draw[cyan,fill=cyan] (0,0) circle (.7ex); Networked Minds Subscale~\cite{biocca2001networked, nowak2003effect}\end{tabular} & ~\cite{wang2020again, dey2017effects} \\
 &  &  &  & & \begin{tabular}[c]{p{4cm}}\tikz\draw[magenta,fill=magenta] (0,0) circle (.7ex); Conversation turn-taking\end{tabular} & ~\cite{smith2018communication} \\ \cline{2-7} 
 & \begin{tabular}[c]{p{1.8cm}}Satisfaction of interaction\end{tabular} &11 & \begin{tabular}[c]{p{3.5cm}}Reinforces social bonds and strength, Increases further engagements\end{tabular} & \begin{tabular}[c]{p{3.5cm}}Personalized content, Adaptive operation difficulty levels, etc.\end{tabular} &  \begin{tabular}[c]{p{4cm}}\tikz\draw[cyan,fill=cyan] (0,0) circle (.7ex); Immersive Experience Questionnaire~\cite{jennett2008measuring}\end{tabular} & ~\cite{drey2022towards, wang2020again} \\ \cline{2-7} 
 & \multirow{2}{*}{\begin{tabular}[c]{p{1.8cm}}Engagement\end{tabular}}&\multirow{2}{*}{15} & \multirow{2}{*}{\begin{tabular}[c]{p{3.5cm}}Sustains interest and participation, Extends the duration and depth of social interactions\end{tabular}}  &\multirow{2}{*}{\begin{tabular}[c]{p{3.5cm}}Immersive spaces, Interactive objects/effects, Shared activities, etc.\end{tabular}}  &  \begin{tabular}[c]{p{4cm}}\tikz\draw[cyan,fill=cyan] (0,0) circle (.7ex); Conversation Engagement Subscales~\cite{garau2003impact}\end{tabular} & ~\cite{wang2020again, li2021evaluating} \\
 &  &  &  &  & \begin{tabular}[c]{p{4cm}}\tikz\draw[magenta,fill=magenta] (0,0) circle (.7ex); Speech duration, Unique word count metric, Duration and Frequency of eye contact, Body or Head orientations\end{tabular} & ~\cite{abdullah2021videoconference, smith2018communication} \\ \bottomrule[2pt]
\end{tabular}%
}
\end{table}

\textbf{Interpersonal experience.}
 This dimension reflects users' perceptions of each other in VR settings, crucial for assessing VR's potential in fostering communication, connection, and relationship development among users. Researchers have detailed this aspect through the following aspects.

 \textit{Mutual awareness}, measured by 17 studies, refers to the shared understanding of others' presence and actions, which leads to more authentic interactions and encourages engagement and responsiveness with each other~\cite{biocca2003toward}. This experience is associated with avatar fidelity and the naturalness of users' gestures reconstruction~\cite{smith2018communication, aseeri2021influence}. The commonly used scales include the co-presence module of Nowak and Biocca questionnaire~\cite{biocca2003toward, nowak2003effect} and the Game Experience Questionnaire~\cite{ijsselsteijn2007characterising}. Sample items from these questionnaires include ``I often felt that my partner and I were sitting together in the same space.''~\cite{li2019measuring}

 \textit{Psychological involvement}, measured by 21 studies, refers to users' degree of cognitive and emotional investment to others, which could be facilitated by emotion-sharing features~\cite{salminen2018bio, dey2017effects} and the activities encouraging personal experiences sharing~\cite{wei2023bridging, baker2020evaluating}. When users are psychologically involved with others, they are more likely to understand others' emotions and develop deeper connections~\cite{nowak2003effect, biocca2003toward}. It is often assessed by measuring participants' perceived closeness, trust, and importance toward others. Studies reported them by collecting self-reported data using the Social Connectedness questionnaire~\cite{van2009social}, the Interpersonal Trust Scale~\cite{rotter1967new}, and so on. Sample questions from these scales include ``I was emotionally close to your partner''~\cite{martikainen2019collaboration}. 
 Studies also tracked `interpersonal distance' as a measurement of psychological involvement. Different interpersonal distances (e.g., intimate, personal, social, or public distances) were employed to assess participants' perceived closeness in VR contexts~\cite{roth2018beyond, buck2019interpersonal, williamson2021proxemics, choudhary2021revisiting, bonsch2018social, choudhary2020virtual, rivu2021friends, sra2018your}. 

 \textit{Affective Interdependence} reflects users' influence and reliance on others' emotional states, moods, and feelings, which was measured by nine studies. This experience could be facilitated by emotion-sharing features~\cite{salminen2018bio, dey2017effects} and the activities required emotional goals~\cite{salminen2018bio}. Studies measured participants' affective interdependence by analyzing their self-reported data from subjective questionnaires, such as the Networked Minds subscales~\cite{biocca2001networked}. The sample questions are ``I was able to feel my partner's emotions''~\cite{wang2020again}, and ``I was influenced by my partner's moods''~\cite{salminen2018bio}. Few studies detected participants' heartbeat sync rate to measure this experience~\cite{dey2017effects}.

\textbf{Interaction experiences.}
 Measuring these experiences could help to understand whether and how the VR design promotes satisfied communication and collaboration experiences. Reviewed studies documented users' interaction experiences by following dimensions.

 \textit{Quality of interaction} refers to the depth, effectiveness, and overall positive experience. In this review, eleven studies assessed participants' quality of interaction through self-reported data from subjective scales such as the Networked Minds subscale~\cite{biocca2001networked, nowak2003effect} or the Game Experience Questionnaire~\cite{ijsselsteijn2007characterising}. Example questions include ``I could fully understand what my partner was talking about''~\cite{li2019measuring}. One study tracked participants' conversation turn-taking (e.g., lower turn frequency reflects the lower efficiency of interaction) to report the quality of interaction~\cite{smith2018communication}.

 \textit{Satisfaction of interaction} refers to the level of contentment or fulfillment experienced during their social interaction, which was measured by eleven studies. Studies showed that personalized content (e.g., objects, spaces)~\cite{mcveigh2019shaping, baker2021school} and skill-appropriated operations~\cite{wei2023bridging} could promote this experience. Studies assessed how well expectations were met, the enjoyability of the interaction, and whether desired outcomes were achieved by using subjective questionnaires based on the Immersive Experience Questionnaire~\cite{jennett2008measuring}, Intrinsic Motivation Inventory~\cite{ryan2000self}, and so on. Sample questions include ``I really enjoyed the time spent with my partner''~\cite{aseeri2021influence}.
 
\textit{Engagement} refers to active involvement, attention, and participation in social interaction, which was measured by 15 studies. Maintaining engagement is crucial for sustaining participation and extends the duration and depth of social interactions~\cite{rajagopalan2015play, herzog2002social}. Eight studies measured engagement through self-reported data from subjective questionnaires based on the Networked Minds~\cite{biocca2001networked}, the Conversation Engagement subscales~\cite{garau2003impact}, and so on. In these assessments, participants reported their attention allocations to different objects (e.g., interaction partner, objects) to indicate the degree of engagement. 
Moreover, twelve studies tracked participants' actual behaviors to assess their engagement. Among these, verbal behaviors were the most commonly used tracking data, such as the communication frequency/turn-taking~\cite{dey2017effects, abdullah2021videoconference, smith2018communication}, the speech duration~\cite{montagud2022towards, sykownik2020like, li2019measuring}, and the unique word count metric (i.e., the number of distinct words spoken)~\cite{aseeri2021influence, smith2018communication, sykownik2020like}.
Additionally, eight studies utilized the duration and frequency of eye contact as indicators of participants' engagement in interpersonal interaction~\cite{roth2018beyond, vinnikov2017gaze, abdullah2021videoconference, kimmel2023let}. Five estimated relative participants' engagement based on their body or head orientations~\cite{wang2020again, dey2017effects, aseeri2021influence, biocca2003toward, kimmel2023let}.


\subsubsection{Experimental Settings} Examining studies' experimental settings helps us understand how researchers simulated and observed participants' social interactions in VR environments, and how they collected valuable data from participants. This section explores task design, data collection methods, and study sites reported in studies. Table~\ref{tab:studysetting} presents the statistical analysis of this dimension.

 \begin{table}[htb!]
\renewcommand{\arraystretch}{1.1}
    \caption{Experimental Settings}    
    \label{tab:studysetting}
    \begin{tabular}{cp{8cm}c}
    \toprule[1pt]
        \textbf{Experimental Settings}  & \textbf{Statistic Analysis (\% of studies)}          \\ \toprule[0.5pt]
        Task design   & Prototype exploration (34.04\%); Collaborative tasks (11.70\%); Competitive tasks (3.19\%); Talking with given topics (11.70\%); Talking freely (9.57\%) \\
         \midrule[0.5pt]
        Data collection   &  Qualitative data:  Interviews (52.13\%); Contextual inquiry (8.51\%); Co-design workshop (4.26\%); Focus Groups (2.13\%); Dairy (2.13\%); Online comments collection (2.13\%) \\ 
        & Quantitative data:  Questionnaire (46.81\%); Behavioral and Biological data (21.28\%); Online surveys (5.32\%)\\
        \midrule[0.5pt]
        Study site     &   Laboratories (92.55\%); Real-world environments (7.45\%) \\
        \bottomrule[1pt]
    \end{tabular}
    
\end{table}

\textbf{Task design.}
Studies employed various experimental tasks in VR to simulate different social environments and trigger user interaction, as detailed in Table~\ref{tab:studysetting}. The largest portion of studies (34.04\%) adopted \textit{prototype exploration} to let participants experience their VR designs in specific social contexts, such as gaming~\cite{sykownik2023vr, dey2017effects, sykownik2020like, choudhary2021revisiting, huard2022cardsvr}, co-watching videos~\cite{rothe2021social, montagud2022towards}, and photo sharing~\cite{li2019measuring}. These activities facilitated focused data collection and analysis of user experiences to inform targeted VR design. Additionally, eleven studies (11.70\%) conducted \textit{collaborative tasks} to trigger more interaction among participants, ranging from negotiation exercises~\cite{smith2018communication, he2020collabovr} to intelligence games~\cite{wienrich2018social, martikainen2019collaboration, fiani2023big}. Three studies (3.19\%) also incorporated \textit{competitive tasks}, such as investment games~\cite{lin2023measuring, george2018trusting} and Whac-A-Mole~\cite{fribourg2018studying}, to examine aspects such as strategy and conflict resolution in competitive contexts.

In addition to interaction tasks, 20 studies incorporated conversational tasks. Among these, eleven studies (11.70\%) prompted participants to \textit{discuss specific topics}, facilitating communication on various subjects from decision-making to casual conversations~\cite{shih2023do, roth2016avatar, baker2019interrogating, baloup2021non, aburumman2022nonverbal, jung2021use, baker2021school, aseeri2021influence, abdullah2021videoconference}, such as opinions on health trends~\cite{fu2023mirror}. Conversely, nine studies (9.57\%) adopted an unstructured approach, allowing \textit{free talks} without predefined tasks or topics to encourage natural interactions and potentially reveal organic social behaviors in VR (e.g., ~\cite{moustafa2018longitudinal, wei2023bridging, kimmel2023let}). This approach offers insights into individuals' natural navigation and communication within virtual spaces.

\textbf{Data collection.} 
In the reviewed studies, the most frequently employed qualitative data collection method was the interview, with 49 studies (52.13\%) using \textit{Interviews} to gather participants' subjective opinions and feedback on their social experiences in VR (e.g., ~\cite{piitulainen2022vibing, freeman2022disturbing, blackwell2019harassment}). Additionally, eight studies (8.51\%) employed \textit{contextual inquiry}, where researchers entered social VR platforms to actively engage in the users' activities while observing both system design~\cite{Tanenbaum2020, Maloney2020, osborne2023being} and users' specific social behaviors~\cite{sabri2023challenges, acena2021my, wienrich2018social, Marcel2019towards, kolesnichenko2019understanding}. This approach provided insights into how users engage in specific social interactions within the naturalistic environments of VR.
Furthermore, four studies (4.26\%) engaged in \textit{co-design} activities with participants, enhancing system design through direct user input in a participatory development process~\cite{lee2022designing, wei2023bridging, baker2019exploring, lee2022understanding}. Two studies (2.13\%) hold \textit{focus group} to discuss specific topics and gather diverse perspectives of participants~\cite{mei2021cakevr, lee2022designing}. \textit{Diaries} were used in two studies (2.13\%), allowing participants to record their daily experiences with the VR systems, thereby providing a longitudinal perspective on user experience~\cite{levordashka2023exploration, abramczuk2023meet}. Additionally, two studies (2.13\%) \textit{analyzed online comments} from social VR users on social media to examine and present authentic user experiences~\cite{zamanifard2019togetherness, zamanifard2023social}.

For the quantitative data, the \textit{subjective questionnaire} was the most common method, used in 44 studies (46.81\%)~\cite{li2021evaluating, dey2017effects, roth2018beyond, salminen2018bio}. Participants provided feedback on their social experiences and the usability of the VR platforms after completing study tasks. Twenty studies (21.28\%) collected \textit{behavioral data and biological signals} from users while interacting with the VR systems, such as measuring interpersonal distances~\cite{roth2018beyond, buck2019interpersonal, williamson2021proxemics}, linguistic features~\cite{smith2018communication, hoeg2021buddy}, and other behavioral indicators~\cite{wang2020again, dey2017effects, aseeri2021influence, biocca2003toward, kimmel2023let}. Lastly, five studies (5.32\%) conducted \textit{online surveys} targeting specific themes among social VR users, providing valuable data on specific user perceptions and interactions within VR environments~\cite{fang2021social, deighan2023social, sykownik2022something, sykownik2021most, lee2022understanding}.

\textbf{Study site.} Among the studies reviewed, only seven studies (7.45\%) were conducted in real-world environments (e.g., participants' homes or workplaces)~\cite{fu2023mirror, moustafa2018longitudinal, levordashka2023exploration, wei2023bridging, abramczuk2023meet, williamson2021proxemics, xu2022social}. In contrast, the majority of user experiments have been conducted in laboratory settings (92.55\%)~\cite{smith2018communication, abdullah2021videoconference, baker2021school}. This raises concerns about the naturalness of the communication behaviors observed under such controlled conditions. To better understand user communication behavior and experience in more natural states, future research should prioritize experiments involving multiple remote users in their typical environments.


\subsubsection{Participant Characteristics} Understanding participant characteristics helps to evaluate the generalizability of study findings across diverse populations, identify potential biases, and guide future research directions. In this section, we present the participant characteristics of 79 studies that involved participants (shown in Table~\ref{tab:participant}), excluding system observational studies~\cite{Tanenbaum2020, Maloney2020, osborne2023being}, those analyzing social media comments related to social VR~\cite{zamanifard2019togetherness, zamanifard2023social}, and large-scale online surveys~\cite{fang2021social, deighan2023social, sykownik2022something, sykownik2021most, lee2022understanding}. 

 \begin{table}[htb!]
\renewcommand{\arraystretch}{1.1}
    \caption{Participant Characteristics}    
    \label{tab:participant}
    \begin{tabular}{cp{8cm}c}
    \toprule[1pt]
        \textbf{Participant Characteristics}  & \textbf{Statistic Analysis (\% of studies)}          \\ \toprule[0.5pt]
        Sample size     &  Mean: 34.19; Median: 27; Sample range: 4 to 210; Interquartile range: 19 to 42.5\\
         \midrule[0.5pt]
        Group Size & 1 participant (16.46\%); 2 participants (44.30\%); 3-5 participants (10.13\%); More than 5 participants (3.80\%) \\
        \midrule[0.5pt]
        Participant Diversity  &  32.97\% studies involving participants from under-represented populations, in terms of age (20.21\%), sexes (17.02\%), physical or mental conditions (6.33\%), and sexual orientations (1.27\%)\\
        \midrule[0.5pt]
        Participant Relationships  & 12.66\% studies involving participants with pre-existing relationships, including teammates/classmates (5.06\%), family (3.80\%), friend (2.53\%), and acquaintances (2.53\%) \\
        \midrule[0.5pt]
        Participants' VR Expertise   & All participants had VR experience (43.03\%); Mixed VR Experience (32.91\%); No Prior VR Experience (11.39\%); Not Reported  (31.65\%)\\
        \bottomrule[1pt]
    \end{tabular}
    
\end{table}

\textbf{Sample size}. The studies had a mean of 34.19 and a median of 27 participants, indicating that while some studies recruited larger groups, the typical study size remained relatively modest. The number of participants in these studies ranged significantly, from 4~\cite{sykownik2020like} to 210~\cite{abdullah2021videoconference}. This range highlights the varied scopes and resources across the studies. The interquartile range of 19 to 42.5 participants suggests that most studies clustered around this range, providing insight into the common scale of participant involvement. Among these participants, 62.69\% are male, 36.05\% are female, and 1.26\% are identifying as other. This revealed a gender imbalance among participants.


\textbf{Group Size}. For our review papers, most studies (44.30\%) involve \textit{two participants} per group in a VR setting to assess dyadic social interactions. 13 studies (16.46\%) introduced only \textit{one participant} per experimental session and employed animated avatars to simulate the presence of other users, such as computer-generated avatars~\cite{vinnikov2017gaze, lin2023measuring, aburumman2022nonverbal, choudhary2021revisiting, latoschik2019not, wang2021shared, yamada2021visual, choudhary2020virtual, bonsch2018social}, researcher participation as avatars~\cite{baloup2021non, shih2023do, fu2023mirror}, or even requiring participants to imagine the presence of other users~\cite{muthukumarana2020touch, jamil2021emotional}. Additionally, eight studies (10.13\%) increased the group size to \textit{3-5 participants}~\cite{abdullah2021videoconference, levordashka2023exploration, baker2019interrogating, baker2021school, baker2021avatar, moustafa2018longitudinal, xu2023designing, roth2018beyond}. This range allows for the exploration of multi-person social interactions in VR, thereby enriching the variety of social scenarios and interactions that can be studied. 
Three studies (3.80\%) have investigated social experiences in VR involving groups \textit{larger than five participants}~\cite{abramczuk2023meet, williamson2021proxemics, xu2022social}. However, these studies did not limit participants to a specific type of device, allowing for both VR headsets and desktop applications. Consequently, a thorough investigation has not been conducted into social VR experiences that exclusively use headsets for larger groups. 

\textbf{Participant Diversity.} Involving participants from under-represented/minority populations in the study is crucial for capturing varied perspectives and fostering inclusive VR social platforms. In this review, 31 studies (32.97\%) report involving participants from under-represented populations. Specifically, 19 studies (20.21\%) included participants who were children~\cite{fiani2023big}, adolescents~\cite{kimmel2023let}, middle-aged~\cite{morris2023don}, or elderly~\cite{baker2019exploring}, 16 studies (17.02\%) included LGBTQ participants~\cite{freeman2022queer, abramczuk2023meet}, five studies (6.33\%) included participants with physical or mental conditions~\cite{maloney2020anonymity, maloney2020complicated}, and one study included participants with less common sexual orientations~\cite{acena2021my}.
These studies revealed disparities in social VR experiences among different user groups, contributing to developing sensitivity and inclusivity in VR design.

\textbf{Participant Relationships.} Among the papers reviewed, ten studies (12.66\%) specifically recruited participants who had pre-existing relationships with each other, such as family members~\cite{wei2023bridging, moustafa2018longitudinal, hoeg2021buddy}, friends~\cite{moustafa2018longitudinal, wang2020again}, teammates/classmate~\cite{levordashka2023exploration, levordashka2023exploration, abramczuk2023meet, xu2022social}, or acquaintances~\cite{li2019measuring, rothe2021social}. Most studies, however, did not involve those who already knew each other to investigate communication patterns or interaction design within established relationships. Since social interaction in real life often occurs between people who are familiar with each other, future research could further explore how different relationships influence communication patterns and interaction design.

\textbf{Participants' VR Expertise and VR training}. Participants' VR expertise is critical, as unfamiliarity with VR operation and novelty effect impact their interaction experiences~\cite{miguel2023countering, huang2020investigating}. Our review reveals a diversity of VR expertise among participants. Among these studies, 34 (43.03\%) invited participants who had experience with VR, primarily to assess their previous experiences with commercial social VR applications (e.g.,~\cite{freeman2022queer, li2023we, morris2023don, lee2022designing}). 26 studies (32.91\%) included a mix of experienced and inexperienced VR users (e.g.,~\cite{wang2020again, shriram2017all, baker2021avatar, fribourg2018studying}), while nine studies (11.39\%) involved participants who had never used VR before (e.g.,~\cite{shih2023do, levordashka2023exploration, roth2016avatar, roth2018beyond}). Additionally, 25 studies (31.65\%) did not report the VR experience of their participants (e.g., ~\cite{wienrich2018social, bonsch2018social, xu2022social, muthukumarana2020touch}). 

Implementing VR training and warm-up sessions before the study could minimize the impacts of unfamiliarity with VR operations and the novelty effect~\cite{wei2023bridging, objectJ}. However, among the reviewed studies, only 24 (30.38\%) reported conducting comprehensive VR training and warm-up sessions to ensure participants were familiar with the VR operations and environment (e.g., ~\cite{wang2020again, smith2018communication, baker2019interrogating, li2021evaluating}). This gap underscores the need for more rigorous preparation in VR studies to ensure consistent and reliable results.

\section{Discussion and Future Directions}

In this section, we first summarize our findings and suggest guidelines for future research to design better VR features for enhancing users' social experiences. Subsequently, based on our results, we highlight the \sout{the overlooked} \rv{underexplored}  areas in the reviewed papers and discuss higher-level research directions that can inform future research in social VR.

\subsection{Design Implications for Social VR} Our findings highlight three VR features crucial for social interaction, acknowledging both positive and negative effects. Building upon these findings, we discuss the design implications for each feature in future social VR applications, aiming to cultivate social VR environments that enhance positive experiences while mitigating negative ones.

Firstly, user self-representation through avatars in VR can significantly influence self-perception and social behavior~\cite{freeman2021body, yee2007proteus}. Customizable avatar appearances enable users to create ideal self-perceptions and foster anonymity, facilitating interaction without fear of judgment~\cite{zamanifard2023social, freeman2022re}. Therefore, future VR designers should provide a broader range of VR identities for users to explore different personas and modes of interaction~\cite{freeman2021body, freeman2022disturbing}.
Additionally, our review shows that avatar fidelity contributes to different social atmospheres. Future designers should consider offering avatars with varying levels of fidelity for different scenarios, such as abstract avatars for informal activities~\cite{wei2023bridging, freeman2021body} and realistic avatars for formal settings~\cite{aseeri2021influence}. 
However, the use of avatars can also diminish users' sense of accountability, potentially leading to negative behaviors such as cyberbullying and misinformation dissemination~\cite{blackwell2019harassment, freeman2022re}. Therefore, VR developers should implement robust mechanisms for detecting and reporting harmful behavior, ensuring a harmonious social experience in social VR applications.

Secondly, VR offers various interaction strategies for users to express themselves, including interactions resembling reality and interactions beyond reality. Consequently, users convey their thoughts and intentions more naturally and effectively in VR compared to other social media platforms, such as phone calls or video chat~\cite{abdullah2021videoconference, brubaker2012focusing, dey2017effects, kratz2014polly}. An increasing number of researchers are exploring enhancing VR experiences to be more realistic or even surpass reality~\cite{delazio2018force, jung2021use, hamdan2019springlets, huard2022cardsvr, roth2019technologies, roth2018beyond}. However, despite its potential, some studies caution that the rich feedback in VR may not always improve social interactions~\cite{li2023we, freeman2022disturbing, blackwell2019harassment}. For instance, realistic haptics could amplify the negative impact of inappropriate behavior, potentially causing harm~\cite{li2023we}. Therefore, future designs should prioritize user comfort while providing immersive interactions. This could involve offering preference settings for intimate interactions and allowing users to customize feedback fidelity to their comfort.

Lastly, we highlight the role of the environment in VR in influencing the social norms and content of interaction patterns~\cite{devito2007interpersonal}. 
The suitable VR environment offers specific communication contexts, which foster shared experiences and promote more active engagement~\cite{mcveigh2019shaping, wei2023bridging, freeman2021hugging}. Therefore, it's important to craft suitable VR environments for supporting specific social contexts. 
Furthermore, research indicates users' desire for personalized VR environments to tailor unique social experiences~\cite{wei2023bridging, mcveigh2019shaping}. Previous studies have explored generating 3D objects or scenes from 2D sketches or text using artificial intelligence-generated content (AIGC)~\cite{li2023generative, ma2018language, he2020collabovr}. We suggest leveraging this technology for VR scene customization, such as generating environments relevant to users' ongoing conversation topics to foster deep discussions, or crafting atmospheres based on users' emotional states during interactions to influence their moods.


\subsection{Future Research Directions}
Based on the materials reviewed, we uncovered several research gaps and proposed potential future research directions, including developing theoretical foundations in studies (based on Sec.~\ref{sec:des}), investigating the effects of VR's environmental factors (based on Sec.~\ref{sec:features}), leveraging VR's positive effects \sout{support} \rv{to support} interpersonal interaction (based on Sec.~\ref{sec:context} and Sec.~\ref{sec:effects}), mitigating VR's negative effects by establishing effective regulatory frameworks (based on Sec.~\ref{sec:effects}), and evolving methodologies for future studies (based on Sec.~\ref{sec:measure}). By pointing out these future directions, we aim to advance the field of VR research and contribute to the development of more effective, inclusive, and ethical VR technologies.

\subsubsection{Developing Theoretical Frameworks for Social VR} 

Across the reviewed studies, we \rv{noticed} a recurring trend similar to other domains in the HCI field~\cite{objectJ, ReplicationsHCI}: a tendency to prioritize novelty over integrating research with established theories. Many papers focus on exploration and descriptive findings, overlooking the importance of anchoring observations within existing theoretical frameworks for comprehensive interpretation. This lack of theoretical grounding can undermine the depth, rigor, reusability, developmental potential, and societal relevance of conclusions. Thus, we advocate for future research to bridge this gap by \textbf{linking their findings to established theories}. This could involve fostering interdisciplinary collaboration and engaging experts from various fields to examine VR applications through multifaceted theoretical lenses. Researchers can enrich their understanding of HCI phenomena by incorporating insights from disciplines like psychology, sociology, and communication studies, promoting diversity and depth in their investigations~\cite{carroll2003hci, roto2021mapping}.

Furthermore, our analysis indicates that while some papers integrate theory, they predominantly draw from non-VR contexts, such as 2D gaming or real-life settings, to inform social VR design. Only a few studies have developed~\cite{li2019measuring} and applied~\cite{wang2021shared, montagud2022towards, wang2020again} frameworks grounded in social VR's unique characteristics, which blend aspects of reality with virtual interactions. The distinct nature of social VR poses challenges for applying theories from other domains, as they may not fully capture or explain the observed social phenomena in VR. Thus, there is a critical need to \textbf{develop theoretical frameworks tailored specifically to the social VR environment}. This effort entails organizing empirical findings into systematic frameworks or methodologies to deepen understanding of VR's applications and impacts across diverse contexts. Establishing such frameworks will be essential for guiding future research and advancing the field of social VR.

\subsubsection{Exploring the Effects of VR's Environmental Factors} 
Researchers have demonstrated that various VR features significantly influence users' social perceptions and behaviors~\cite{dey2017effects, smith2018communication}. However, our review shows that most studies have primarily focused on exploring the effects \sout{brought by} \rv{of} avatar appearance and interaction strategies in VR. There is a noticeable gap in research regarding \textbf{exploring the impact of VR's environmental factors on social interaction}. In this review, only two studies examined the effects of VR's environmental factors (e.g., space size) on users' social behavior~\cite{sra2018your, williamson2021proxemics}. Therefore, we encourage future researchers to explore the effects of other VR environmental factors (including style, decor, orderliness, organization, and openness) on users' social behavior. In the realm of environmental psychology, numerous studies \sout{investigate} \rv{have investigated} the impact of physical spaces on user behavior in the real world~\cite{mehta2007lively, hipp2019effect}, such as the tendency of people to break the rules in the presence of chaotic visual information~\cite{VRPhysical}. 
However, it remains unclear whether these findings can be applied in a VR setting to shape user perceptions and social behavior. Consequently, we propose that future research  \sout{could} \rv{should} explore how \sout{the} environmental factors in VR can influence users' social behavior. This exploration could provide valuable insights into optimizing VR environments for more active and harmonious social interactions.

\subsubsection{Leveraging the Benefits of Social VR to Support Various Interpersonal Interactions}
Based on the current landscape of social VR research discussed in Sec~\ref{sec:context}, we found that most reviewed papers primarily explore how VR supports users' social needs in public or group settings, with fewer studies investigating how VR can facilitate interpersonal interactions. \sout{Combined with the benefits of} \rv{Building upon VR's advantages for} interpersonal interactions, we propose several research directions that warrant exploration in the future.

First, future research could explore \textbf{using VR to encourage equitable and unbiased inter-group interactions}. Potential applications include facilitating cross-cultural, inter-generational, and inter-racial social interactions. Stereotyping and prejudice among different user groups often hinder effective communication~\cite{williams2013intergenerational}. VR could potentially mitigate these biases and promote inclusivity by blurring identity differences through digital avatars~\cite{moustafa2018longitudinal, wei2023bridging}. Additionally, users can embody avatars of different identities to experience diverse social perspectives. The sense of embodiment in VR encourages users to think from others' perspectives, fostering empathy and understanding~\cite{smith2018communication, li2019measuring, salminen2018bio}. Therefore, future research should explore how VR can facilitate social interactions across various ages, genders, races, and cultural groups to dismantle preconceived notions and achieve inclusive interaction. For example, \sout{letting} \rv{allowing} younger users \sout{embody} \rv{to embody} an older avatar may \sout{arouse} \rv{invoke} their empathy toward the elderly, while the elderly may perceive younger users as their equals~\cite{wei2023bridging}. This could provide an equal communication opportunity for them to \sout{get} \rv{gain} a deeper mutual understanding.

Second, we encourage future research to \textbf{investigate VR for better support of intimate relationships}. Our review shows that participants appreciated VR's ability to provide a sense of togetherness and companionship, which are essential for nurturing close relationships by fostering intimacy and overcoming geographical distance~\cite{tu2002relationship, biocca2003toward}. However, only a few studies have focused on participants with close relationships~\cite{wei2023bridging, zamanifard2019togetherness}. Future research should explore how VR can support communication and interaction design tailored to different types of close relationships, enhancing their effectiveness and satisfaction.

Finally, we advocate exploring VR to \textbf{support different stages of relationship development}, such as conflicts and disagreements. Current studies mainly focus on using VR to enhance mutual understanding and emotional exchange. However, relationships are dynamic, and conflicts and disagreements can arise during social interaction~\cite{devito2007interpersonal}. VR allows users to switch perspectives by embodying \sout{other's} \rv{others'} avatars and offers a relaxed communication atmosphere for calm discussions. This \sout{potentially resolves} \rv{could help resolve} misunderstandings and foster reconciliation. 

\subsubsection{Mitigating VR Negative Effects by Establishing Effective Regulatory Frameworks Tailored for Social VR} 
While previous research has highlighted negative social experiences in VR, such as harassment and privacy concerns, only a few studies \sout{explored} \rv{have explored} the strategies that social VR platforms can implement to mitigate these negative effects~\cite{schulenberg2023towards}. 
Although several commercial social VR applications have implemented guidelines to manage user behavior to avoid negative social experiences, these guidelines are often rudimentary and lack practical applicability. For example, VRChat's community guidelines state that individual users are responsible for reporting inappropriate or harassing behavior. However, the ambiguity surrounding what constitutes ``inappropriate or harassing behavior'' leaves crucial information obscured~\cite{blackwell2019harassment}. Additionally, researchers have noted that the regulatory mechanisms and policies effective on other platforms cannot be directly applied to social VR because users' social interactions and experiences in social VR are notably different, featuring more embodied interactions~\cite{abdullah2021videoconference, baker2021avatar}. For instance, a close interpersonal distance, acceptable in computer games, can be perceived as intrusive in VR. 

Consequently, we propose future research should explore \sout{the effective way of} \rv{effective methods for} \textbf{establishing a dedicated community to develop regulatory frameworks tailored for social VR} to foster a harmonious communication environment. This includes addressing the following questions: What behaviors are deemed unacceptable in social VR? How do these behaviors impact user interactions? How can these undesirable behaviors be detected, and what would constitute reasonable penalties for such actions? These issues necessitate extensive discussion and policy-making within the community~\cite{schulenberg2023towards, blackwell2019harassment}.
To establish a more efficient and unbiased regulatory system in social VR, \sout{a} \rv{one} previous study also suggests incorporating user-human-AI collaboration into this process~\cite{schulenberg2023towards}. This approach involves users acting as overseers to ensure impartiality and timely updates, the community (human) as policy-makers to maintain authority, and AI as enforcers to ensure efficiency. This tripartite system would collaboratively uphold the regulatory framework, allowing it to evolve with societal changes and adapt to different contexts, thereby shaping a harmonious and inclusive social VR environment.

With the advancement of VR technology, the amount of time we spend in VR is likely to increase. Consequently, it is crucial to clearly define responsibilities and obligations in VR to build a humane and cohesive society. This could significantly enhance users' social experiences in social VR and increase their willingness to engage with the platform. Therefore, we call for the creation of a specialized social VR community to develop actionable regulatory frameworks to foster a harmonious VR social environment.

\subsubsection{Methodological Evolution of Studies} In terms of methodology evolution, several crucial directions emerge. One important direction is to \textbf{propose a reference and standardized task to assess social experiences in VR}. Whittaker et al. have highlighted the general absence of standardized tasks within HCI~\cite{areferencetaskagenda}. In our review, we also observed a diversity of tasks used in studies aimed at guiding users' social interactions in VR, including prototype exploration~\cite{mei2021cakevr}, collaborative tasks~\cite{smith2018communication}, competitive tasks~\cite{lin2023measuring}, and conversational tasks~\cite{wei2023bridging}. While each of these tasks effectively facilitates social interaction in VR, they only provide insights from specific social contexts. However, users' social behaviors vary across different social contexts~\cite{devito2007interpersonal}. Therefore, relying solely on a single social task for users fails to yield generalized results applicable to various social scenarios. We recommend future studies  \sout{to design} \rv{design} standardized social tasks to test VR designs' impact and user experiences across diverse social situations. This approach helps to concentrate on crucial aspects in the field, share metrics and datasets, and advance theory~\cite{areferencetaskagenda}.
 
Secondly, we advocate that future studies should \textbf{conduct ecologically valid user studies} to assess VR designs for social interaction. Currently, the majority of user experiments are characterized by controlled laboratory settings (92.55\% of studies), relatively short duration (averaging less than 60 minutes of VR exposure), and limited VR training (69.62\% did not report VR training and warm-up sessions). Unfortunately, these factors compromise the authenticity and practical applicability of experimental outcomes~\cite{morgado2017scale}. To mitigate these limitations, future research should prioritize ecologically valid user experiments to gain a deeper understanding of user experiences in VR-based social interactions. This entails several key components: studies conducted \textit{outside laboratory settings}, \textit{longer-term studies}, and \textit{adequate VR training sessions}. Among these, longer-term studies and adequate VR training sessions allow researchers to explore the long-term consequences of VR interactions, mitigating the influence of novelty effects associated with VR usage~\cite{wei2023bridging, montagud2022towards, li2019measuring}. Conducting studies outside the laboratory setting enables researchers to observe participants' engagement with VR designs in authentic, relaxed social atmospheres, uncovering the genuine benefits VR offers for socialization. These insights are pivotal for developing VR applications that resonate with users' social needs and preferences.


Thirdly, exploring VR social experiences within \textbf{larger group size} is important. While many commonplace social scenarios, such as group meetings or parties, involve more than five users engaging simultaneously~\cite{devito2007interpersonal}, our review found a lack of studies investigating users' social experiences in ``immersive VR settings'' with groups of more than five participants. Engaging a larger group size in studies presents opportunities to observe intricate group dynamics, evolving social norms, and the complexities of forming enduring relationships within VR social environments. Hence, future research should investigate potential challenges and requirements that social VR users may face when interacting within large groups.

Finally, we advocate for future research to \textbf{include more participants from under-represented populations} to reflect and address the needs of all users. In this review, only 32.97\% of studies reported involving participants from under-represented populations. Furthermore, future research should strive to \textbf{achieve gender balance} in their participant samples. Our results reveal a significant gender imbalance, with 62.69\% of participants being male. Previous studies have revealed the disparities in social VR experiences among different user groups. For example, females \sout{felt more sensitive to} \rv{were more affected by} sexual harassment in VR~\cite{femalesensetive}, disabled users felt that avatars could prevent differential treatment in VR~\cite{maloney2020anonymity}, and non-cisgender users faced potential harassment due to avatar-voice mismatches~\cite{freeman2022disturbing, blackwell2019harassment}. Therefore, it is essential to include diverse participant groups and achieve gender balance in studies to understand social needs from various perspectives and to promote the accessibility and inclusivity of VR environments.

\section{Limitation and Future Work}

In this section, we acknowledge some limitations of our study.
Firstly, our selection \sout{was exclusively} \rv{focused exclusively} from the HCI field, potentially limiting our focus to user experience and VR design and possibly overlooking broader psychological, sociological, and cultural dimensions. To achieve a more comprehensive understanding, we recommend conducting a systematic literature review that samples from various fields to provide a wider range of perspectives. 
Additionally, our review utilized a limited set of keywords for paper selection. Although we expanded our search using Google Scholar to minimize omissions, we cannot ensure that our collection includes every study related to users' social interaction in VR. 
Nevertheless, our goal was to offer a comprehensive summary and analysis of key trends and findings, rather than an exhaustive catalog of all relevant research. Therefore, while there might be some gaps due to these omissions, we believe they do not significantly detract from our review's overall validity and conclusions.

Second, we gathered papers that met our two criteria as our final paper set. These criteria \rv{helped} us \rv{select} the papers that offer sufficient insights for addressing our RQs. Although certain papers \sout{include} \rv{involving} VR studies for social interaction, such as~\cite{qiao2022how, pena2020model, orts2016holoportation}, were not included because they did not meet our two key criteria, we acknowledge the significance of these papers. We propose that future research investigates such papers to explore other pertinent RQs, such as identifying the benefits and challenges of using VR in educational contexts.

Furthermore, the subjective nature of our analysis method could introduce bias. The coding process may be influenced by the researchers' interpretations of themes and concepts, potentially leading to subjective bias~\cite{cooper2019handbook, cooper1982scientific}. Similarly, affinity diagramming, used for organizing and grouping diverse research findings, could result in oversimplification or misclassification of complex ideas due to its inherent limitations~\cite{beyer1999contextual}. To mitigate these issues, we ensured the reliability of codes by iteratively discussing and revising them to resolve conflicts in our weekly meetings among all co-authors. 

\section{Conclusion}

VR is increasingly used as a social platform where users can interact and connect with others in immersive virtual worlds. Despite its growing usage, there is a lack of comprehensive understanding of how VR is concretely being used for socialization and its promising effects. To address this gap, a literature review of 94 papers in the HCI field was conducted using the PRISMA method. Our findings suggest that VR influences social interactions through self-representation that affects self-perception, various interaction strategies for direct communication, and interaction environments that scaffold social behaviors. While VR offers benefits like promoting relaxation, enhancing engagement, fostering intimacy, and improving accessibility, it also has drawbacks, including intensifying harassment experiences and amplifying privacy concerns. We summarized the measurement employed in reviewed studies to measure users' social experiences in VR, including evaluation dimensions, experimental settings, and participant characteristics. Based on these results, we discuss and point out several research directions that need to be explored in the future.

\bibliographystyle{ACM-Reference-Format}
\renewcommand{\bibpreamble}{References marked with *  are in the set of reviewed papers.}
\bibliography{sample-base}

\appendix


\begin{table}[]
\renewcommand{\arraystretch}{1.3}
\caption{Comprehensive List of Evaluation Dimensions and Measurement Tools Used in Existing Literature for Social Experiences in VR. \textcolor{cyan}{Blue} dots indicate self-report scales, while \textcolor{magenta}{red} dots signify tracking data.} 
\label{allmetric}
\resizebox{12cm}{!}{%
\begin{tabular}{l|l|l}
\toprule[2pt]
 & \multicolumn{1}{c|}{\textbf{Evaluation Dimensions}} & \multicolumn{1}{c}{\textbf{Scales and Tracking Data}} \\ \hline
\multirow{16}{*}{\textbf{\begin{tabular}[c]{@{}l@{}}Intrapersonal\\ Experiences\end{tabular}}} & \multirow{2}{*}{Embodiment} & \tikz\draw[cyan,fill=cyan] (0,0) circle (.5ex); The illusion of virtual body ownership (IVBO)\cite{roth2017alpha}(be applied by~\cite{latoschik2017effect}) \\
 &   & \tikz\draw[cyan,fill=cyan] (0,0) circle (.5ex); Avatar embodiment questionnaire~\cite{gonzalez2018avatar}(be applied by~\cite{fribourg2018studying}) \\ \cline{2-3} 
 & \multirow{5}{*}{Immersion} & \tikz\draw[cyan,fill=cyan] (0,0) circle (.5ex); Slater-Usoh-Steed (SUS) Questionnaire~\cite{usoh2000using} (be applied by~\cite{jung2021use, choudhary2021revisiting, bonsch2018social, drey2022towards, sra2018your, wang2020again, li2021evaluating, li2019measuring, montagud2022towards}) \\
 &  & \tikz\draw[cyan,fill=cyan] (0,0) circle (.5ex); Witmer and Singer Presence (WS) Questionnaire~\cite{witmer1998measuring} (be applied by~\cite{huard2022cardsvr, baker2021avatar, latoschik2019not, drey2022towards, wang2020again, li2021evaluating, li2019measuring, montagud2022towards} \\
 &  & \tikz\draw[cyan,fill=cyan] (0,0) circle (.5ex); Igroup Presence Questionnaire (IPQ)~\cite{schubert2001experience}(be applied by~\cite{sykownik2020experience, wang2021shared, wienrich2018social, rothe2021social, drey2022towards, wang2020again, li2021evaluating, li2019measuring, montagud2022towards, rzayev2020s}) \\
 &  & \tikz\draw[cyan,fill=cyan] (0,0) circle (.5ex); Nowak and Biocca (NB) Questionnaire~\cite{nowak2003effect}(be applied by~\cite{latoschik2017effect, roth2016avatar, roth2018beyond}) \\
 &  & \tikz\draw[cyan,fill=cyan] (0,0) circle (.5ex); Spatial Presence Experience Scale (SPES)~\cite{hartmann2015spatial}(be applied by~\cite{van2023feelings}) \\ \cline{2-3} 
 & \multirow{9}{*}{Affective states}  & \tikz\draw[cyan,fill=cyan] (0,0) circle (.5ex); Pictorial Mood Reporting Instrument(PMRI)~\cite{vastenburg2011pmri}(be applied by~\cite{wang2020again, li2019measuring, wang2021shared}) \\
 & & \tikz\draw[cyan,fill=cyan] (0,0) circle (.5ex); Positive and Negative Affect Schedule (PANAS)~\cite{watson1988development, mackinnon1999short}(be applied by~\cite{dey2017effects, wienrich2018social, afifi2022using}) \\
 & & \tikz\draw[cyan,fill=cyan] (0,0) circle (.5ex); Discrete Emotions Questionnaire (DEQ)~\cite{harmon2016discrete}(be applied by~\cite{sykownik2020experience}) \\
 &  & \tikz\draw[cyan,fill=cyan] (0,0) circle (.5ex); Geneva Emotion Wheel~\cite{scherer2005emotions}(be applied by~\cite{moustafa2018longitudinal}) \\
 &  & \tikz\draw[cyan,fill=cyan] (0,0) circle (.5ex); Game Experience Questionnaire (GEQ)~\cite{ijsselsteijn2007characterising}(be applied by~\cite{wienrich2018social}) \\
 &  & \tikz\draw[cyan,fill=cyan] (0,0) circle (.5ex); Social Presence Survey (SPS)~\cite{bailenson2001equilibrium}(be applied by~\cite{bonsch2018social}) \\
 &  & \tikz\draw[cyan,fill=cyan] (0,0) circle (.5ex); Word Pairs Survey~\cite{short1976social}(be applied by~\cite{smith2018communication}) \\
 &  & \tikz\draw[magenta,fill=magenta] (0,0) circle (.5ex); Valence and arousal of their speech~\cite{li2019measuring} \\
 &  & \tikz\draw[magenta,fill=magenta] (0,0) circle (.5ex); Duration of their laughter~\cite{sykownik2020like} \\ \hline
\multirow{25}{*}{\textbf{\begin{tabular}[c]{@{}l@{}}Interpersonal\\ Experiences\end{tabular}}} & \multirow{7}{*}{Mutual awareness} & \tikz\draw[cyan,fill=cyan] (0,0) circle (.5ex); \begin{tabular}[c]{p{10cm}}the co-presence module of Nowak and Biocca questionnaire~\cite{biocca2003toward, nowak2003effect}(be applied by~\cite{smith2018communication, vinnikov2017gaze, latoschik2017effect, aseeri2021influence, afifi2022using, wang2020again, li2021evaluating, li2019measuring, montagud2022towards, roth2016avatar, roth2018beyond})\end{tabular} \\
 &  & \tikz\draw[cyan,fill=cyan] (0,0) circle (.5ex); Game Experience Questionnaire (GEQ)~\cite{ijsselsteijn2007characterising}(be applied by~\cite{wienrich2018social, sra2018your}) \\
 &  & \tikz\draw[cyan,fill=cyan] (0,0) circle (.5ex); Bailenson questionnaire~\cite{bailenson2003interpersonal}(be applied by~\cite{jung2021use, george2018trusting}) \\
 &  & \tikz\draw[cyan,fill=cyan] (0,0) circle (.5ex); Multimodal Presence Scale~\cite{makransky2017development}(be applied by~\cite{van2023feelings}) \\
 &  & \tikz\draw[cyan,fill=cyan] (0,0) circle (.5ex); Poeschl and Doering questionnaire~\cite{poeschl2015measuring}(be applied by~\cite{rzayev2020s}) \\
 &  & \tikz\draw[cyan,fill=cyan] (0,0) circle (.5ex); Steed questionnaire~\cite{steed1999leadership}(be applied by~\cite{jung2021use}) \\
 &  & \tikz\draw[cyan,fill=cyan] (0,0) circle (.5ex); Garau and Slater questionnaire~\cite{garau2005responses}(be applied by~\cite{choudhary2021revisiting}) \\ \cline{2-3} 
 & \multirow{14}{*}{Psychological involvement} & \tikz\draw[cyan,fill=cyan] (0,0) circle (.5ex); Social Connectedness questionnaire~\cite{van2009social}(be applied by~\cite{wang2020again, li2021evaluating, li2019measuring, montagud2022towards}) \\
 &  & \tikz\draw[cyan,fill=cyan] (0,0) circle (.5ex); Social Connection questionnaire~\cite{tarr2018synchrony}(be applied by~\cite{martikainen2019collaboration}) \\
 &  & \tikz\draw[cyan,fill=cyan] (0,0) circle (.5ex); social connectedness scale from Carroll's wore~\cite{carroll2017conceptualization}(be applied by~\cite{sykownik2023vr}) \\
 &  & \tikz\draw[cyan,fill=cyan] (0,0) circle (.5ex); Affective Benefits and Costs(ABC) questionnaire~\cite{ijsselsteijn2009measuring}(be applied by~\cite{rothe2021social}) \\
 &  & \tikz\draw[cyan,fill=cyan] (0,0) circle (.5ex); Unidimensional Relationship Closeness Scale~\cite{dibble2012unidimensional}(be applied by~\cite{afifi2022using}) \\
 &  & \tikz\draw[cyan,fill=cyan] (0,0) circle (.5ex); Togetherness questionnaire~\cite{basdogan2000experimental}(be applied by~\cite{sra2018your}) \\
 &  & \tikz\draw[cyan,fill=cyan] (0,0) circle (.5ex); Subjective Closeness Index~\cite{gachter2015measuring}(be applied by~\cite{roth2018beyond}) \\
 &  & \tikz\draw[cyan,fill=cyan] (0,0) circle (.5ex); the Rapport questionnaire form Gratch's work~\cite{gratch2015negotiation}(be applied by~\cite{latoschik2017effect}) \\
 &  & \tikz\draw[cyan,fill=cyan] (0,0) circle (.5ex); Competitive and Cooperative Presence in Gaming Questionnaire~\cite{hudson2014measuring}(be applied by~\cite{wienrich2018social}) \\
 &  & \tikz\draw[cyan,fill=cyan] (0,0) circle (.5ex); Multidimensional Scale~\cite{zimet1988multidimensional}(be applied by~\cite{van2023feelings}) \\
 &  & \tikz\draw[cyan,fill=cyan] (0,0) circle (.5ex); Trust questionnaire~\cite{chun1974dimensionality}(be applied by~\cite{latoschik2017effect}) \\
 &  & \tikz\draw[cyan,fill=cyan] (0,0) circle (.5ex); Interpersonal Trust Scale~\cite{rotter1967new}(be applied by~\cite{george2018trusting}) \\
 &  & \tikz\draw[cyan,fill=cyan] (0,0) circle (.5ex); Socio-Economic Panel Scale (SOEP-trust)~\cite{naef2009measuring}(be applied by~\cite{george2018trusting}) \\
 &  & \tikz\draw[magenta,fill=magenta] (0,0) circle (.5ex); Interpersonal distances~\cite{roth2018beyond, buck2019interpersonal, williamson2021proxemics, choudhary2021revisiting, bonsch2018social, choudhary2020virtual, rivu2021friends, sra2018your} \\ \cline{2-3} 
 & \multirow{4}{*}{Affective Interdependence} & \tikz\draw[cyan,fill=cyan] (0,0) circle (.5ex); Networked Minds subscales~\cite{biocca2001networked}(be applied by~\cite{roth2016avatar, wang2020again, li2021evaluating, li2019measuring, montagud2022towards}) \\
 &  & \tikz\draw[cyan,fill=cyan] (0,0) circle (.5ex); Networked Minds Social Presence Measure~\cite{harms2004internal}(be applied by~\cite{salminen2018bio}) \\
 &  & \tikz\draw[cyan,fill=cyan] (0,0) circle (.5ex); Game Experience Questionnaire (GEQ)~\cite{ijsselsteijn2007characterising}(be applied by~\cite{wienrich2018social}) \\
 &  & \tikz\draw[cyan,fill=cyan] (0,0) circle (.5ex); Interpersonal Reactivity Index~\cite{davis1980multidimensional}(be applied by~\cite{martikainen2019collaboration}) \\ \hline
\multirow{18}{*}{\textbf{\begin{tabular}[c]{@{}l@{}}Interaction\\ Experiences\end{tabular}}} & \multirow{5}{*}{Quality of interaction} & \tikz\draw[cyan,fill=cyan] (0,0) circle (.5ex); Networked Minds subscale~\cite{biocca2001networked, nowak2003effect}(be applied by~\cite{wang2020again, dey2017effects, li2021evaluating, li2019measuring, montagud2022towards} \\
 &  & \tikz\draw[cyan,fill=cyan] (0,0) circle (.5ex); Game Experience Questionnaire (GEQ)~\cite{ijsselsteijn2007characterising}(be applied by~\cite{wienrich2018social}) \\
 &  & \tikz\draw[cyan,fill=cyan] (0,0) circle (.5ex); Networked Minds Social Presence Measure~\cite{harms2004internal}(be applied by~\cite{salminen2018bio}) \\
 &  & \tikz\draw[magenta,fill=magenta] (0,0) circle (.5ex); Conversation turn-taking~\cite{smith2018communication} \\
 &  & \tikz\draw[magenta,fill=magenta] (0,0) circle (.5ex); Speech duration~\cite{hoeg2021buddy} \\ \cline{2-3} 
 & \multirow{5}{*}{Satisfaction of interaction} & \tikz\draw[cyan,fill=cyan] (0,0) circle (.5ex); Immersive Experience Questionnaire (IEQ)~\cite{jennett2008measuring}(be applied by~\cite{drey2022towards, wang2020again, li2021evaluating, li2019measuring, montagud2022towards}) \\
 &  & \tikz\draw[cyan,fill=cyan] (0,0) circle (.5ex); Intrinsic Motivation Inventory (IMI)~\cite{ryan2000self}(be applied by~\cite{hoeg2021buddy}) \\
 &  & \tikz\draw[cyan,fill=cyan] (0,0) circle (.5ex); key Components of User Experience~\cite{minge2014modulare}(be applied by~\cite{wienrich2018social}) \\
 &  & \tikz\draw[cyan,fill=cyan] (0,0) circle (.5ex); Interpersonal Communication Satisfaction Scale~\cite{hecht1978conceptualization}(be applied by~\cite{aseeri2021influence}) \\
 &  & \tikz\draw[cyan,fill=cyan] (0,0) circle (.5ex); Player Experience Inventory (PXI) questionnaire~\cite{abeele2020development}(be applied by~\cite{drey2022towards}) \\ \cline{2-3} 
 & \multirow{8}{*}{Engagement} & \tikz\draw[cyan,fill=cyan] (0,0) circle (.5ex); Networked Minds~\cite{biocca2001networked} \\
 &  & \tikz\draw[cyan,fill=cyan] (0,0) circle (.5ex); Conversation Engagement subscales~\cite{garau2003impact} \\
 &  & \tikz\draw[cyan,fill=cyan] (0,0) circle (.5ex); attention to behavioral cues questionnaire~\cite{roth2018some} \\
 &  & \tikz\draw[magenta,fill=magenta] (0,0) circle (.5ex); Communication frequency/turn-taking~\cite{dey2017effects, hoeg2021buddy, abdullah2021videoconference, smith2018communication} \\
 &  & \tikz\draw[magenta,fill=magenta] (0,0) circle (.5ex); Speech duration~\cite{montagud2022towards, sykownik2020like, li2019measuring, hoeg2021buddy} \\
 &  & \tikz\draw[magenta,fill=magenta] (0,0) circle (.5ex); Unique word count metric~\cite{aseeri2021influence, smith2018communication, sykownik2020like} \\
 &  & \tikz\draw[magenta,fill=magenta] (0,0) circle (.5ex); Duration and frequency of eye contact~\cite{roth2018beyond, vinnikov2017gaze, abdullah2021videoconference, kimmel2023let} \\
 &  & \tikz\draw[magenta,fill=magenta] (0,0) circle (.5ex); Body or head orientations~\cite{wang2020again, dey2017effects, aseeri2021influence, biocca2003toward, kimmel2023let} \\ \hline
\end{tabular}%
}
\end{table}

\end{document}